\documentclass{aa}  

\usepackage{graphicx, color}
\usepackage{amsmath}
\usepackage{multirow,bigdelim}
\usepackage{calc}
\usepackage{threeparttable}
\usepackage{amsmath}
\usepackage{array}
\usepackage{tabularx}
\usepackage{caption}
\usepackage[font=small]{caption}
\usepackage[labelformat = empty,position=top]{subcaption}
\usepackage[export]{adjustbox}
\usepackage{mathptmx}
\usepackage{grffile}
\usepackage{pdflscape}
\usepackage[normalem]{ulem}
\usepackage{txfonts}
\definecolor{orange}{RGB}{255,165,0}

\definecolor{dgreen}{rgb}{0., 0.7, 0.}

\definecolor{tiffany}{RGB}{79, 166, 158}

 \newcommand{\Msun}{\mbox{${M}_{\odot}$}}
 \newcommand{\Zsun}{\mbox{${Z}_{\odot}$}}
\newcommand{\Mi}{\mbox{$M_{\rm i}$}}
\newcommand{\Zi}{\mbox{$Z_{\rm i}$}}
\newcommand{\Zcr}{\mbox{$Z_{\rm cr}$}}

\newcommand{\Zcrse}{\mbox{$Z_{\rm crse}$}}
\newcommand{\Mcum}{\mbox{$M_{\rm cum}$}}
\newcommand{\Nvmo}{\mbox{$N_{\rm VMO}$}}
\newcommand{\Mqot}{\mbox{$M4{\rm \omega}3$}}
\newcommand{\Mqoq}{\mbox{$M4{\rm \omega}4$}}
\newcommand{\Mcot}{\mbox{$M5{\rm \omega}3$}}
\newcommand{\Mcoq}{\mbox{$M5{\rm \omega}4$}}
\newcommand{\Msot}{\mbox{$M6{\rm \omega}3$}}
\newcommand{\Msoq}{\mbox{$M6{\rm \omega}4$}}
\newcommand{\Ohl}{\mbox{\rm 12+log(O/H)}}
\newcommand{\MUP}{\mbox{${M_{\rm UP}}$}}
\newcommand{\xUP}{\mbox{${x_{\rm UP}}$}}

\newcommand{\Mchar}{\mbox{${M}_{\rm char}$}}
\newcommand{\Mhe}{\mbox{${M}_{\rm He}$}}

\begin{document}

   \title{The impact of very massive stars on the chemical evolution of extremely metal-poor galaxies}

   \subtitle{}

   \author{S. Goswami
           \inst{1,4}
          \and
           L. Silva
           \inst{2,3}
           \and
           A. Bressan
           \inst{1}
           \and
           V. Grisoni
           \inst{1,2,11}
           \and           
           G. Costa  
          \inst{5,8,9}
           \and           
           P. Marigo  
          \inst{5}
          \and
          G. L. Granato
          \inst{2,3,6}
           \and
          A. Lapi 
          \inst{1,3,7} 
           \and
          M. Spera
          \inst{1,5,8,10} 
           }

   \institute{
        SISSA, Via Bonomea 265, I-34136 Trieste, Italy\\
        \email{sgoswami@sissa.it; sbressan@sissa.it; 
        } 
        \and
        INAF-OATs, Via G. B. Tiepolo 11, I-34143 Trieste, Italy\\
        \email{laura.silva@inaf.it}
        \and
        IFPU - Institute for fundamental physics of the Universe, Via Beirut 2, 34014 Trieste, Italy
        \and
        CENTRA, Instituto Superior Técnico, Av. Rovisco Pais 1, 1049-001 Lisboa
        \and
        Dipartimento di Fisica e Astronomia, Universit\`a degli studi di Padova,
        Vicolo Osservatorio 3, Padova, Italy
        \and
        Instituto de Astronomia Teorica y Experimental (IATE), Consejo Nacional de Investigaciones Cientificas y Tecnicas de la Republica Argentina (CONICET)
        \and
        IRA-INAF, Via Gobetti 101, 40129 Bologna, Italy
        \and
        INFN, Sezione di Padova, Via Marzolo 8, I--35131, Padova, Italy
        \and
        INAF - Osservatorio Astronomico di Padova, Vicolo dell'Osservatorio 5, I-35122, Padova, Italy
        \and
        CIERA and Department of Physics \& Astronomy, Northwestern University, Evanston, IL 60208, USA
        \and 
        Dipartimento di Fisica e Astronomia, Università degli Studi di Bologna, Via Gobetti 93/2, I-40129 Bologna, Italy
        \
        }

   \date{}


\abstract
{In recent observations of extremely metal-poor low-mass starburst galaxies, almost solar Fe/O ratios are reported, despite N/O ratios consistent with the low metallicity.}  
{We investigate if the peculiar Fe/O ratios can be a distinctive signature of an early enrichment produced by very massive objects dying as Pair-Instability Supernovae (PISN).}  
{We run chemical evolution models with yields that account for the contribution by PISN. We use both the recent non-rotating stellar yields from \citet{Goswami2020}, and new yields from rotating very massive stars calculated on purpose in this work. We also search for the best initial mass function (IMF) that is able to reproduce the observations.} 
{We can reproduce the observations by adopting a bi-modal IMF and by including an initial burst of rotating very massive stars. Only with a burst of very massive stars can we reproduce the almost solar Fe/O ratios at the estimated young ages. We also confirm that rotation is absolutely needed to concomitantly reproduce the observed N/O ratios.}  
{These results stress the importance of very massive stars in galactic chemical evolution studies and strongly support a top-heavy initial mass function in the very early evolutionary stages of metal poor starburst galaxies.} 

\keywords{Chemical ejecta, stellar evolution, nucleosynthesis, chemical evolution of galaxies.}

\maketitle

%

\section{Introduction}






The study of local metal-poor galaxies can greatly help to shed light on early galaxy evolution since they are expected to represent a significant baryon component in the early universe.\\
Extremely metal-poor galaxies (EMPGs), with metallicities less than 12+log(O/H)=7.69 \citep{Kunth2000,Izotov2012,Isobe2020}, being 12+log(O/H)=8.69 the solar metallicity \citep{Asplund2009}, are a particular example of this class that has been discovered both in the local universe and at  high redshifts.  Local EMPGs have low stellar masses (log($M_{\star}/M_{\odot}$) $\sim$6-9) and high specific star formation rates (i.e.\ SFR per unit stellar mass, sSFR$\sim$10-100 Gyr$^{-1}$) \citep{Izotov1998,Thuan2005,Pusti2005,Izotov2009,Izotov2018,Izotov2019,Skillman2013,Hirsch2016,Hsyu2017}. They are considered as the analogs of high-z galaxies, with similar metallicities and low stellar masses discovered 
at redshift z$\sim$2-3 \citep{Christensen2012a,Christensen2012b,Vanzella2017} and z$\sim$6-7 \citep{Stark2017,Mainali2017}.
\\Recently, a new EMPGs survey called "Extremely Metal-Poor Representatives Explored by the Subaru Survey"  \citep[EMPRESS,][]{Kojima2020} has been initiated with wide-field optical imaging data obtained in Subaru/Hyper Suprime-Cam \citep[HSC;][]{Miya2018} and Subaru Strategic Program \citep[HSC/SSP;][]{Aihara2018}. In particular, \cite{Kojima2021} presented element abundance ratios of local EMPGs from EMPRESS and the literature. They found that neon- and argon-to-oxygen abundance ratios (Ne/O, Ar/O) are similar to those of known local dwarf galaxies, and that the nitrogen-to-oxygen abundance ratios are lower than $\sim$ 20\% of the solar N/O value, in agreement with the low oxygen abundance. Regarding the iron-to-oxygen abundance ratios Fe/O, they found that their metal-poor galaxies show a decreasing trend in Fe/O ratio as metallicity increases, similar to what was found in the star-forming sample of \cite{Izotov2006}, but with two representative EMPGs with exceptionally high Fe/O ratios at the low metallicity end.
\\In \cite{Kojima2021}, three scenarios which might explain the observed Fe/O ratios of their EMPGs are described, as summarised in the following.
\\The first scenario is based on the preferential dust depletion of iron \citep{Rodri2005,Izotov2006}.
In this case, it is assumed that the gas-phase Fe/O abundance ratios of the EMPGs decrease with metallicity because Fe is depleted into dust more efficiently than O; this depletion becomes important at higher metallicities, when dust production is more efficient. However, \cite{Kojima2021} do not find evidence that galaxies with a larger metallicity have a larger colour excess, i.e. that they are dust richer. Hence, the Fe/O decrease of their sample should not be due to dust depletion, and thus they exclude this scenario.
\\The second scenario invokes the presence of metal enrichment and gas dilution due to inflow. In this case, it is assumed that metal-poor galaxies formed from metal-enriched gas having solar metallicity and Fe/O ratio. Then, if primordial gas falls onto metal-enriched galaxies, the metallicity (inferred by the oxygen abundance O/H) decreases, whereas the Fe/O ratio does not change. This scenario might explain the almost solar Fe/O ratios, but one would expect an almost solar N/O ratio as well; the two peculiar EMPGs with high Fe/O ratios, have low N/O ratios (lower than 20\% of the solar value), at variance with what would have been expected by this second scenario, that should therefore be ruled out.
\\Finally, the third scenario refers to the contribution of super massive stars. \cite{Ohkubo06}  showed that stars with masses larger than $\sim 300$ \Msun\ can produce a large amount of iron during supernova (SN) explosion, and hence \cite{Kojima2021} suggested that this contribution of iron could result in the high Fe/O ratios of the two peculiar EMPGs.
In this case, the N/O ratio would not be affected by SN explosion, as required by observations \citep{1999ApJS..125..439I,Ohkubo06}.

While the latter scenario by \cite{Kojima2021} is appealing, other possible solutions could be conceived, especially those that take into account the detailed chemical evolution of the elements in the environmental conditions of such galaxies. There is a large literature on the chemical evolution of galaxies and in particular on the mechanisms that may give rise to the different abundance ratios observed in the Milky Way(MW) and in nearby galaxies \citep[e.g.][and reference therein]{matteucci2021A&ARv}.
\\Concerning metal-poor galaxies, relevant progress 
to explain the observed spread in abundance ratios of selected elements like He, N, O has been made with the introduction of differential outflows, i.e.\ with the assumption that the ejecta of SNII may leave the potential well of the galaxies more efficiently than the ejecta of intermediate- and low-mass stars   \citep[e.g.][]{pilyugin1993A&A,marconi1994MNRAS}, or even that different SNII ejecta may escape with different efficiencies via the so-called selective winds \citep[e.g.][]{MacLow1999ApJ,recchietal2001MNRAS,Fujita2004ApJ,ott2005MNRAS, recchietal2007A&A,recchietal2008A&A,salvadorieial2008MNRAS,Salvadori2014MNRAS,salvadori2019MNRAS}.
In particular, \citet{recchietal2008A&A} showed that variations of abundance ratios, including [O/Fe],
could indeed be obtained by assuming different relative efficiencies in the outflows of the O and Fe ejected from supernovae, but that such differences are unlikely to happen in the early phases of a starburst \citep{recchietal2004A&A}, which are the typical condition of the EMPG galaxies we are analysing here.
\\Thus, in the following, we will reconsider the scenario favoured by \citet{Kojima2021}, with the help of new yields that include the effects of Pulsational Pair Instability Supernovae (PPISN) and Pair Instability Supernovae  \citep[PISN,][]{Goswami2020}.\\

It is worth noting that PISN yields have been already included in the analysis of the chemical evolution of galaxies by other authors, however with a motivated constraint that they should not be able to form  above  a certain threshold metallicity,  Z$_{cr}\lesssim 10^{-4} Z_{\odot}$  \citep{Schneider_raf2002ApJ,salvadorieial2008MNRAS}. The reason for this choice is that detailed studies of the fragmentation process show that the characteristic fragment mass drops dramatically if the effects of the first formed dust are added to metal line cooling. In this case, typical fragments of solar or subsolar masses are obtained for any metallicity Z $>$ Z$_{cr}$ \citep{Schneider2006MNRAS}. Instead, if dust cooling does not operate, the same models predict typical final fragments of hundred solar masses up to a metallicity Z$_{cr}\lesssim 10^{-2} Z_{\odot}$.\\
On the observational side, there has been compelling spectroscopic evidence \citep{Crowtheretal2016MNRAS} of very massive single stars with initial masses up to about 300 \Msun\ populating young starburst regions in the Large Magellanic Cloud (LMC). The spectroscopic VLT-FLAMES Tarantula Survey, which is monitoring the 30 Dor region in the LMC \citep{Schneideretal2018A&A,Schneider2018Sci}, has revealed the presence of single stars with ages between 1 and 6 Myr, and a general mass distribution well populated up to 200 \Msun. Furthermore, the derived overall IMF of stars more massive than 15 \Msun\ has a slope of $x=0.9$ ($x=1.35$ being the Salpeter one), with the individual region around NGC2070 showing $x=0.65$.   
The LMC has a typical metallicity Z$\sim$0.3 \Zsun\ \citep{costaetal2019MNRAS}, much larger than  the lower value of Z$_{cr}$ predicted by theoretical models. Some of these stars are already evolved and, given their high surface helium content, they will likely be affected by the electron-pair creation instability process in their following evolution, and end up as PPISN and/or PISN supernovae. 
From the stellar evolution side, only large mass-loss rates may prevent the formation of PISN by such massive stars. With the current models, the PISN fate is possible up to Z$\sim$ 0.5 \Zsun\ \citep{Kozyreva2014a, Langer2012, Costaetal2020}, above which mass-loss rates are indeed so strong to prevent this SN channel. \\
To overcome this tension between theory and observations, it has been argued that very massive objects (VMO) might arise from early merging during their evolution, e.g.\ as members of interacting binaries. However, the young age of some of the most massive stars in the 30 Dor region does not favour a scenario in which all of them are the results of merger events \citep{Crowtheretal2016MNRAS, Schneider2018}.\\


The effects of the PISN phase in the abundance ratios of EMPG galaxies can be tested using the recent models for non-rotating massive and very massive stars provided by \cite{Goswami2020}, which include the contribution from PPISN and PISN. 
A peculiarity of the ejecta of PISN is the production of large amounts of Fe and O, in proportions that are strongly dependent on the He core mass at the end of the central H burning phase  \citep{Heger_Woosley2002y, Kozyreva2014a,Takahashi2018ApJ857}. 
Indeed chemical evolution models that include PISN yields may possess a very early phase with a low [O/Fe] and then rapidly evolve into the domain of the $\alpha$-enhanced regime  \citep{Goswami2020}.
\\Based on these findings, we will test here whether the high Fe/O together with the low N/O abundance ratios of EMPGs can be both explained using the yields of very massive stars or if additional ingredients are required.\\


The paper is structured as follows. In Section \ref{chemod} we describe the simple chemical evolution model used in this work. In particular, to include the PISN yields, we 
need to shift the upper mass limit of the IMF in the domain of very massive objects.
In this section, we also discuss the PISN yields with particular emphasis on an update of the \cite{Goswami2020} tables that concern the effects of stellar rotation. 
In Section \ref{data}, we present the observational data that we wish to analyse. In Section \ref{evoabu}, we compare the observations to the results of suitable chemical evolution models stressing the derived evolutionary constraints. 
In Section \ref{cheIMF}, we present our choice for a bi-modal IMF that, together with a model that includes a burst of rotating very massive objects, is able to well reproduce the observations.
Finally, in Section \ref{conclu}, we summarise our conclusions.



\section{Chemical evolution model}\label{chemod}

The chemical evolution code used in this work is \texttt{CHE-EVO} \citep{Silva1998}. This is a one-zone open chemical evolution model, which follows the time evolution of the gas abundances of elements, including infall of primordial gas. This code has been used in several contexts to provide the input star formation and metallicity histories to interpret the spectrophotometric evolution of both normal and starburst galaxies 
\citep[e.g.][]{Vega2008,Fontanot2009,Silva2011,Lofaro2013,Lofaro2015,Hunt2019}.
The basic equation used in this code can be written as follows:
\begin{equation}
\dot{M}_{{\rm g},j}= \dot{M}_{{\rm g},j}^{\rm SF}  + \dot{M}_{{\rm g},j}^{\rm Inf} +  \dot{M}_{{\rm g},j}^{\rm FB} ,
\label{eq_chem_ev}
\end{equation}
where, for the element \textit{j}, the first term on the right,   $\dot{M}_{{\rm g},j}^{\rm SF}$ represents the rate of gas consumption by star formation, $\dot{M}_{{\rm g},j}^{\rm Inf}$ corresponds to the infall rate of pristine material and $\dot{M}_{{\rm g},j}^{\rm FB}$ refers to the rate of gas return to the interstellar medium (ISM) by dying stars.
The latter term also includes the contribution of type Ia supernovae (SNIa), whose rate is adjusted with the parameter
${A_{\rm SNIa}}$ that corresponds to the fraction of binaries with system mass between  3\Msun\ and 16\Msun, with the right properties to give rise to SNIa \citep{Matteucci1986}.
\\We use the Schmidt-Kennicutt law \citep{Kenni1998} to model the star formation rate (SFR):
\begin{equation}
\psi(t)= \nu \,  M_{\rm g}(t)^{k} 
\label{eq_SF_law}
\end{equation}
where $\nu$  is the efficiency of star formation, $M_{\rm g}$ is the mass of the gas and $k$ is the exponent of the star formation law. 
\\The gas infall law is assumed to be exponential \citep[e.g.][]{Grisoni_2017,Grisoni2018}, with an e-folding timescale $\tau_{\rm inf}$.
\\The IMF can be written as follows:
\begin{equation}
\phi(M_{\rm{i}}) =  \frac{dn}{d\log(M_{\rm{i}})} \propto M_{\rm{i}}^{-x}.
\label{imfkroupa}
\end{equation}
%
%

We use a Kroupa-like three-slope power law IMF with $x=0.3$ for $0.1 \le \Mi/\Msun \le\, 0.5$, $x=1.2$ for $0.5 \le \Mi/\Msun \le\, 1$, and we change the upper mass limit ($M_{UP}$) and the slope of the upper IMF (\xUP) in the high mass domain  \citep{marks_imf_2012MNRAS} to explore the importance of very massive stars in the chemical evolution of EMPGs.
%
%
%
\\The input parameters of the chemical evolution models considered in this work are summarised in Table \ref{parameter}. In the following, we give details on the stellar and nucleosynthesis models providing our adopted ejecta for the chemical evolution.

\begin{table}
\footnotesize
\caption{Input parameters of the selected chemical evolution and yield models. MTW is our set of stellar ejecta described in \citep{Goswami2020}. These models are shown in Fig.\ \ref{pisn1}.
}
\label{parameter}
\centering
\tiny
\begin{tabular}{|c|c|c|c|c|c|c|c|l|}
\hline
\multicolumn{1}{|c|}{{Name}} &
\multicolumn{4}{|c|}{{Chemical evolution}} &
\multicolumn{2}{|c|}{{IMF}}&\\
\hline
\multicolumn{1}{|c|}{{}} &
\multicolumn{1}{|c|}{${\nu}$ } &
\multicolumn{1}{|c|}{${k}$} &
\multicolumn{1}{|c|}{${\tau_{\rm inf}}$ } &
\multicolumn{1}{|c|}{${A_{\rm SNIa}}$} &
\multicolumn{1}{|c|}{$M_{UP}$}&
\multicolumn{1}{|c|}{$\xUP$} &
\multicolumn{1}{|l|}{{Yields}}\\
\multicolumn{1}{|c|}{{}} &
\multicolumn{1}{|c|}{(Gyr$^{-1}$)} &
\multicolumn{1}{|c|}{{}} &
\multicolumn{1}{|c|}{(Gyr)} &
\multicolumn{1}{|c|}{{}} &
\multicolumn{1}{|c|}{($M_\odot$)}&
\multicolumn{1}{|c|}{{}} &
\multicolumn{1}{|l|}{{}}\\
\hline
M1 &  0.8 & 1.0 & 6.0& 0.04 &100&1.5& MTW \\  
\hline
M2 &  0.8 & 1.0 & 6.0& 0.04 &100&0.6& MTW \\   
\hline
M3 &  0.3& 1.0 & 1.0& 0.04 &300&0.6& MTW \\  
\hline
\end{tabular}
\end{table}

\subsection{Stellar yields}\label{stellaryields}


For low- and intermediate-mass stars ($M_i <$ 8 $M_{\rm \odot}$), we distinguish between single stars and binary systems that give rise to SNe Ia \citep{Matteucci1986}. 

For single stars with initial masses $M_i <$ 6 $M_{\rm \odot}$ we consider yields from calibrated Asymptotic Giant Branch (AGB) models by \cite{Marigo2020NatAs}. 

For $6\Msun < M_i < 8 \Msun $, we consider super-AGB yields from \cite{Ritter_etal18}. 

For SNIa 
we use the yields provided by \cite{1999ApJS..125..439I}.

For $8$ $M_{\rm \odot} \leq M_i \leq$ $350$ $M_{\rm \odot}$, we adopt the yield compilation by \cite{Goswami2020} for massive stars and very massive objects. 
They include stellar wind ejecta, based on the non-rotating \texttt{PARSEC} models \citep{Bressan2012},
and explosive ejecta carefully extracted from explosive models of electron capture SN (ECSN) \citep{Wanajo_etal_09}, core collapse supernova (CCSN) \citep{Limongi&Chieffi2004}, 
PPISN \citep{Woosley_Heger2002y, Chen2014, yoshida2016, Woosley2017ApJ}, and 
PISN \citep{Heger_Woosley2002y, hegerfryer2003}. 
The yields have been calculated for $Z_{\rm i}$= 0.0001, 0.001, 0.006, 0.02. Though the ejecta of PISN are taken from the zero metallicity models by \cite{Heger_Woosley2002y}, they are very similar to those computed for initial metallicities $Z_{\rm i}$ = 0.001, by  \cite{Kozyreva2014a}.

The effect of PISNe is apparent for  $\Zi \leq $0.001, with substantial production of O, Mg, Si, S, Ar, Ca, Ti and Fe, due to C and O ignition within a collapsing core. For more details, we refer the reader to \cite{Goswami2020}. 

\begin{figure} 
\centering
\resizebox{1.0\hsize}{!}{\includegraphics[angle=0]{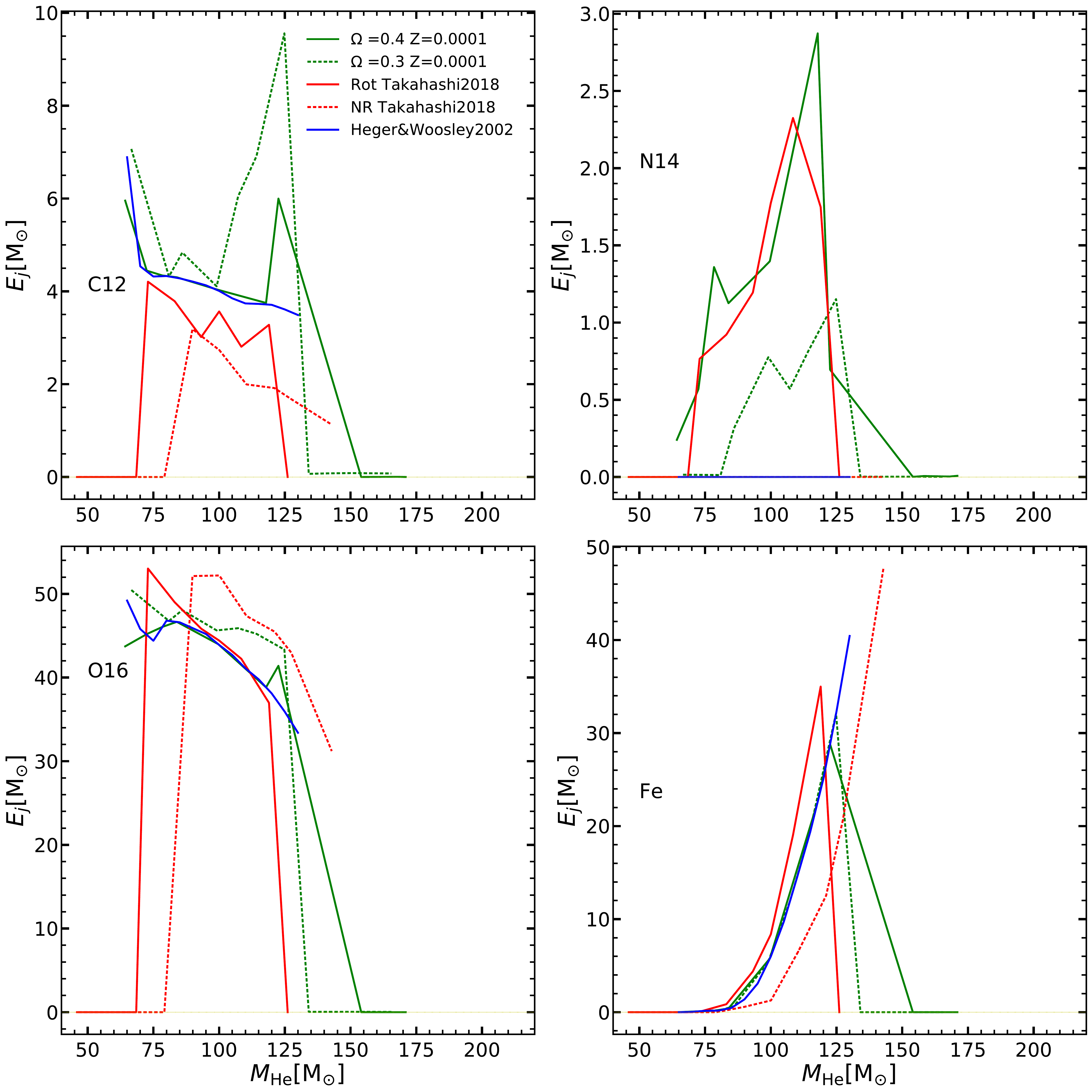}}
\caption{C12, N14, O16 and Fe PISN yields from Z=0.0001 rotational models of this work (green, solid and dotted for 2 rotation rate values) based on \cite{Heger_Woosley2002y} pure He non-rotating models (blue). 
In red, rotating (solid) and non-rotating (dotted) yields by \cite{Takahashi2018ApJ857}. 
Note that N14 yields from non rotating models are negligible.} \label{HW2002} 
\end{figure}

Concerning the concomitant low N/O ratios observed in these EMPGs, it is worth noting that nitrogen production by massive stars is strongly affected by rotation \citep{Limongi_etal18}. In non-rotating massive star models, nitrogen ejecta are much lower than those of rotating models because of extra mixing towards the external layers produced by rotation, and subsequent dispersion by stellar winds \citep{Limongi_etal18,Grisoni2021}.  
This is true also for the N ejecta by PISN, as it can be evaluated by comparing the new ejecta of very massive stars with and without rotation  by \citet{Takahashi2018ApJ857} in Figure \ref{HW2002}.
Even in the case of PISN, nitrogen is essentially produced in the pre-supernova evolution.\\

To fully test the effect of PISN in both O and N production we have computed new yields with the latest PARSEC pre-supernova models with rotation from 100 \Msun\ to 350 \Msun\, up to the beginning of the PISN instability process, as described in  \citet{Costaetal2020}. We have then extracted the explosive yields for $Z_{\rm i}$ = 0.0001, 0.001 and 0.006 from the models by \cite{Heger_Woosley2002y} using the He core mass as the only parameter, as in \cite{Goswami2020}. For the massive stars that do not enter the PISN channel (including the most massive stars with Z$_i$=0.02), we have adopted the yields from \cite{Limongi_etal18}.
A comparison of stellar ejecta from PISN models is shown in Figure \ref{HW2002}.
The ejecta of selected elements from non-rotating PISN models \citep{Heger_Woosley2002y} are shown in blue, while the corresponding ones obtained after matching the PARSEC rotating models are shown in green and, finally the ones for rotating and non-rotating models computed by \citet{Takahashi2018ApJ857} are shown in red, solid and dotted respectively.
By comparing the first two sets we appreciate that the contribution of the pre-SN evolution, i.e. the excess of the green lines above the blue lines, are significant only for the light elements. In particular, we see that nitrogen production from the explosion is negligible with respect to the one arising from the pre-SN evolution.
There are some differences with respect to  \citet{Takahashi2018ApJ857} ejecta, that is quite evident for nitrogen, silicon and magnesium. For the former element, the reason must reside in the different assumptions for the pre-SN models. For the latter elements, the differences must reside in the explosion mechanism. In any case for the other two elements relevant for this paper, oxygen and iron, the different models provide results that are in fair agreement.   


\section{Observational data}\label{data}

The observational data considered in this work are taken from the "Extremely Metal-Poor Representatives Explored by the Subaru Survey" \citep[EMPRESS,][]{Kojima2020}, which provides a database of sources based on the wide-field deep imaging of the Hyper Suprime-Cam Subaru Strategic Program  (HSC-SSP) combined with wide-field shallow data of the SDSS.
From this large program, \cite{Kojima2020} selected a local ($z \lesssim 0.03$) sample of EMPGs, i.e.\ of small galaxies with properties similar to those of high-z galaxies in the early star formation phase 
(low M$_\star$, high sSFR, low metallicity, and young stellar ages) through a machine-learning classifier trained with model templates of galaxies, stars, and QSOs to distinguish EMPGs from other types of objects. 

For 10 of the selected candidates, they conducted spectroscopy follow-ups to infer their detailed physical quantities. Their 10 analysed sources are confirmed as EMPGs,
due to active SF as witnessed by strong emission lines with restframe $H_\beta$ EWs $> 100 \AA$, stellar masses of $\log(M_\star/M_\odot) = 5.0 - 7.1$,  high specific star formation rates of $\sim300$ Gyr$^{-1}$,
young stellar ages $\sim$ 50-100 Myr, and low metallicities. In particular, 3 of these sources strictly satisfy their EMPG criterion of 12+log(O/H) $< 7.69$,  i.e. $Z< 0.1 Z_\odot$, with HSC J1631+4426 showing the lowest metallicity value ever reported, 12+log(O/H) $= 6.90$, (i.e.\ $Z/Z_\odot = 0.016$). The other sources still show very low metallicities ($\sim 0.1 - 0.6$  Z$_\odot$).

For 9 of these selected objects, \cite{Kojima2021} provided also elemental abundance ratios (Fe/O, N/O, Ne/O, Ar/O), and completed their sample with another EMPG, J0811+4730 \citep{Izotov} from the literature, which has the second-lowest reported metallicity (0.0019 Z$_\odot$).
This is the same sample of 10 local EMPGs we consider in our analysis, with the two extreme EMPGs (HSC J1631+4426 and J0811+4730) requiring specific modelling in order to interpret their abundance ratios.
In particular, the ionizing radiation of half of the sampled objects, including these lowest metallicity galaxies, has been found by \cite{Kojima2021} to show both high He II $\lambda$4686/H$_\beta$ ratios (>1/100) together with the high H$_\beta$ EW ($>100 \AA$).
They ascribe these large values to hard radiation from VMOs, which is also their preferred explanation for the exceptionally high Fe/O ratios for some of their objects, as summarised in the Introduction.




In Fig.~\ref{pisn1} we show the Fe/O (left panel) and N/O (right panel) ratios versus the metallicity 12+log(O/H) for these 10 EMPGs (empty and filled circles).
In the left panel, a sample of Galactic thick- and thin-disc stars \citep[blue and magenta dots,][]{Bensbyetal14}, and low metallicity stars \citep[grey dots,][]{Cayrel2004} are shown for comparison.  
In the right panel, the error bars as provided by \cite{Kojima2020} are of the size of the circles plotted in the figure. The light gray dots correspond to local galaxies from \cite{Izotov2006}; the dark gray dots are assembled from \cite{Pettini2002} and \cite{Pettini2008}, which are for extragalactic H II regions and high-redshift DLA systems.

\section {Evolutionary constraints from the abundance ratios}\label{evoabu}

In this section we will test a number of different chemical evolution parameters to understand which evolution scenario is able to produce the high Fe/O ratios and, at the same time, to keep low N/O ratios, as observed in EMPGs 3, 10 and partly 6, that are the three galaxies selected as strictly belonging to the EMPG class.

\begin{figure*} 
\centering
\resizebox{0.45\hsize}{!}{\includegraphics[angle=0]{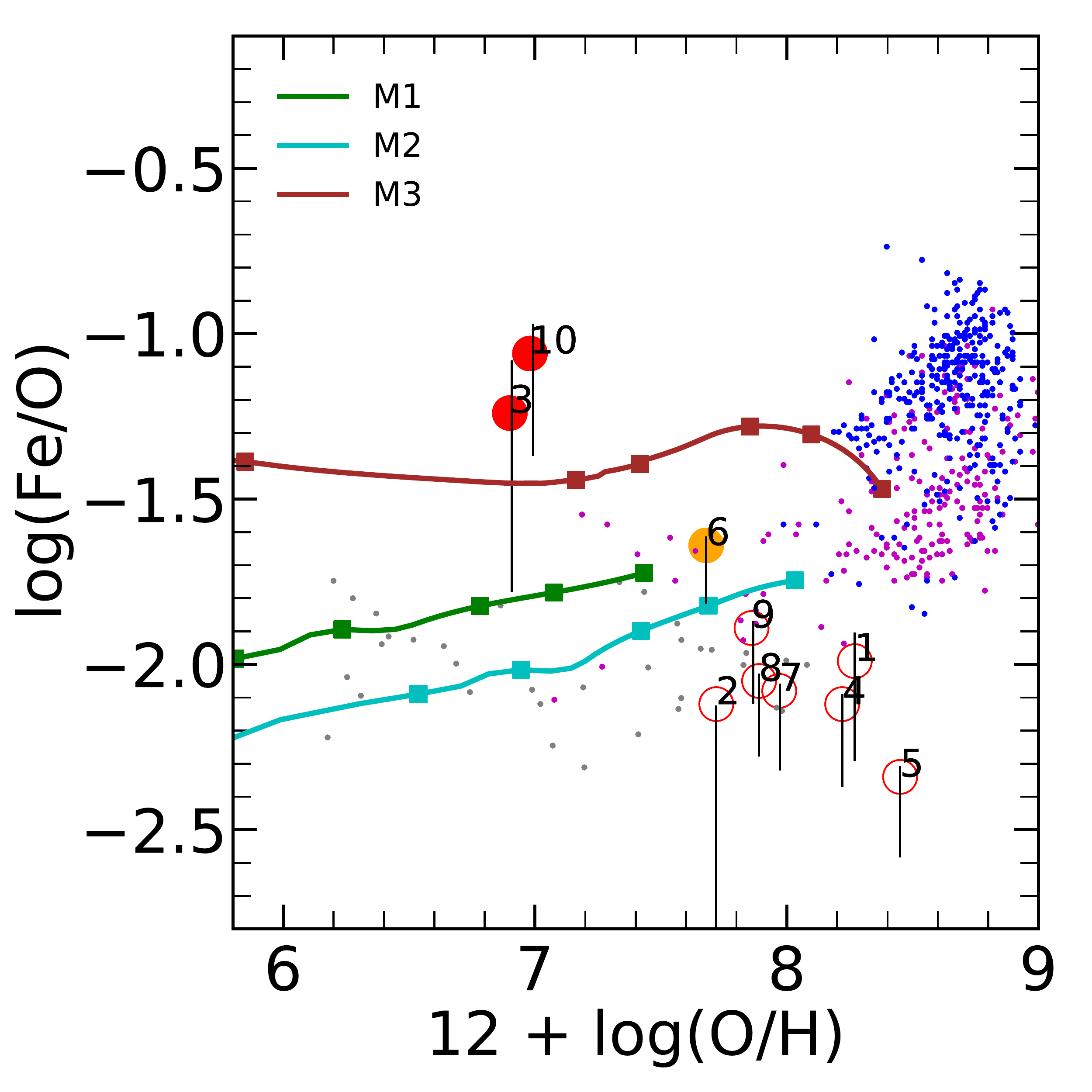}}
\resizebox{0.45\hsize}{!}{\includegraphics[angle=0]{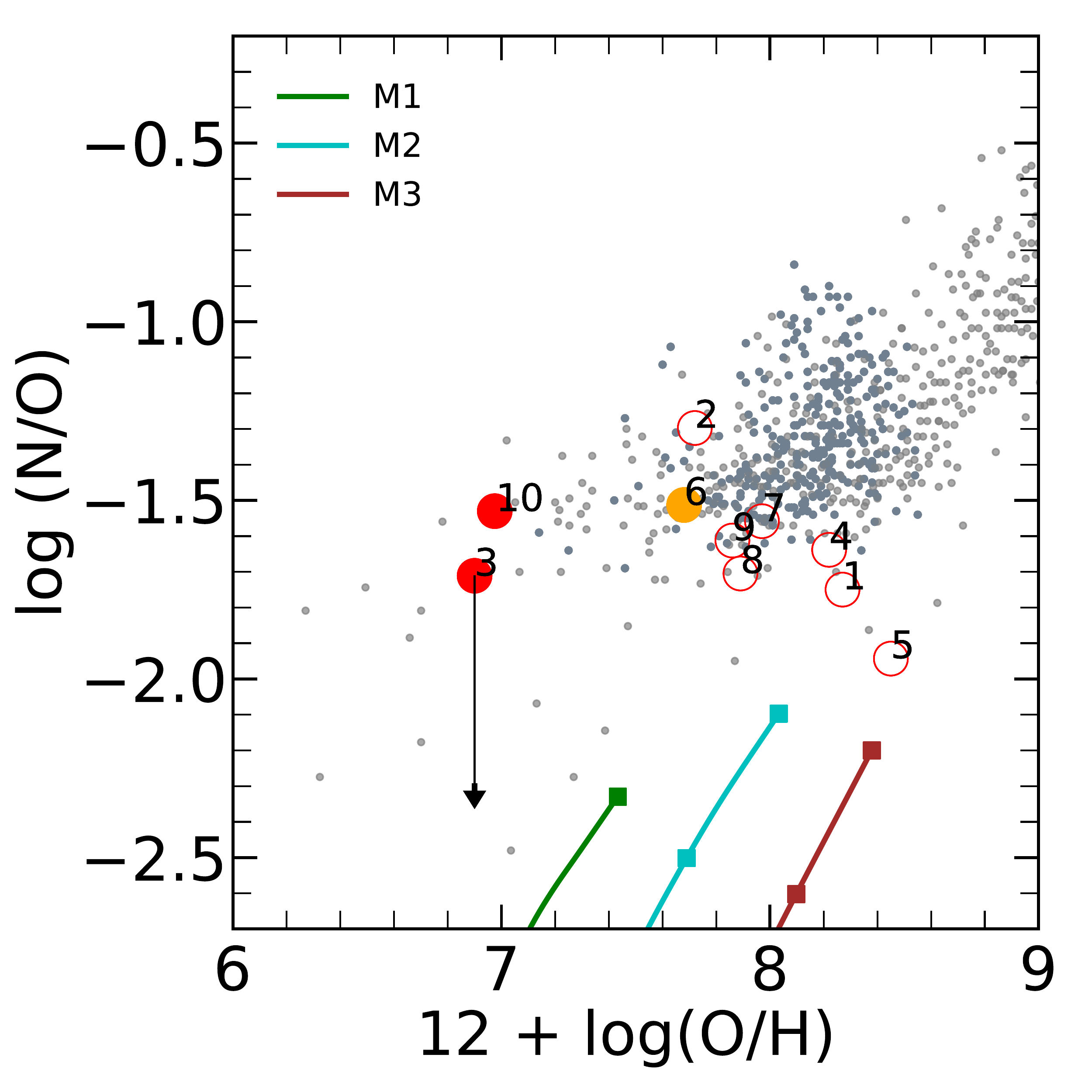}}
\caption{Fe/O ratio (left panel) and N/O ratio (right panel) as a function of the metallicity, 12+log(O/H). EMPGs from \cite{Kojima2021} are shown as numbered circles from 1-9, Object 10 is the EMPG from \cite{Izotov}. The lowest metallicity EMPGs are  are highlighted as filled circles (red for objects 3 and 10, and orange for the intermediate object 6).
In the left panel, blue and magenta dots represent respectively thin and thick disc stars from \citet{Bensbyetal14}, and gray dots are for low metallicity stars from \cite{Cayrel2004}.
In the right panel, light gray dots correspond to local galaxies from \cite{Izotov2006}, dark gray dots are assembled from \cite{Pettini2002} and \cite{Pettini2008}, which are for extragalactic H II regions and high-redshift DLA systems.
Solid line with squares show the evolution of the abundance ratios for the models of Table~\ref{parameter}. 
The six squares mark from left to right  model ages at 5, 20, 30 , 60, 100 and 200 Myr, respectively, the latter being larger than the estimated upper limit by \cite{Kojima2021}. See text for details.
 }
\label{pisn1} 
\end{figure*}

Their peculiar Fe/O abundance ratio,  with respect to other EMPGs of the sample and Galactic disc stars, is illustrated in the left panel of Fig.~\ref{pisn1}. 
To better highlight the peculiarity of these three EMPG galaxies we begin with showing a reference model, M1, whose chemical evolution parameters are within the range of those adopted to reproduce the chemical evolution of the stars in the disc of the MW \citep{Grisoni_2017,Grisoni2018,Grisoni2019,Grisoni2020a,Spitoni2021,Goswami2020}. Model M1 has a \cite{1993MNRAS.262..545K} type IMF with a slope  $\xUP = 1.5$ in the high mass range and an upper mass limit of 100 \Msun, as shown in Table~\ref{parameter}.\\

We see from Fig.~\ref{pisn1}, left panel, that this model is able to reproduce fairly well the location of the thin and thick disc stars of the MW \citep{Bensbyetal14} and also of the tail of halo stars at lower metallicity \citep{Cayrel2004}. We stress that this model is only used as a reference, because the different MW components, being characterized by significantly different metallicity and age distributions, are actually reproduced by different sets of chemical evolution parameters \citep{Goswami2020}. 
From this figure, we may appreciate the peculiarity of EMPG~3 and EMPG~10. Their Fe/O values are clearly well above the region occupied by the other low metallicity objects that, in this region, are almost all below model M1. EMPG~3 has a large error bar that makes it only marginally compatible with model M1. Instead, EMPG~6 could be well fitted by model M1 because, with 12+(O/H)$\sim$7.6, it falls in between the tail of low metallicity stars \citep{Cayrel2004,Bensbyetal14}
that are reproduced by model M1. However if we consider also the temporal evolution of model M1 we immediately see that this match is only apparent. To highlight this point we have marked  six ages with superimposed squares along the M1 path: 5, 20, 30, 60, 100 and 200 Myr, from left to right respectively. 
Model M1 approaches the location of EMPG~6 at $\sim 0.5$~Gyr. This age is about one order of magnitude larger than the maximum age value estimated for EMPGs from their spectro-photometric and nebular emission properties, as described in Section \ref{data} \citep{Kojima2020}.\\
With model M1 we also checked the effect of maximizing the Fe production, with respect to O, by setting  ASNIa=0.95. In this model, not shown here for sake of clarity, the Fe/O ratio reaches at maximum a value close to object 6. Thus, even by increasing the SNIa fraction to almost the maximum allowed value, it is not possible to fit EMPGs 3 and 10. This is because the SNIa enhanced model reaches the O/H values observed for these galaxies in about 100~Myr, a timescale that is too short to allow the evolution of the bulk of intermediate and low mass binary systems, that are the main Fe producers for the selected IMF. We thus confirm that by using an IMF with a canonical \MUP\, like the \cite{1993MNRAS.262..545K}, we cannot reproduce the two EMPGs with the highest Fe/O ratios. \\

The run of the N/O ratio is plotted in the right panel of  Fig.~\ref{pisn1}.
As can be seen from this figure, EMPG 3, 10 and 6 do not show peculiar abundance ratios with respect to other low metallicity objects. Nevertheless, model M1 is not able to reproduce the observed N/O ratios, as in the case of the stars of our galaxy, as  discussed in \citet{Goswami2020}. This is due to a lack of enough N production from non-rotating massive stars. Recently it has been shown that one of the main effects of including rotation in the model of massive stars is that of increasing the N production and indeed only models with rotation may be able to provide the primary N required to fit low metallicity  stars and extragalactic H II regions, or high-redshift DLA systems  \citep{Pettini2002,Pettini2008,Goswami_Prantzos2000,Grisoni2021}. We examine N production below.

\subsection{The Fe/O ratio}

We check whether a different IMF could better reproduce the observed Fe/O ratios of  EMPG~3, ~10 and ~6. With a bottom heavier IMF the resulting metal enrichment would be slower. This would not only exacerbate the age discrepancy already found for EMPG~6, but would likely increase this tension also in the case of EMPG~3 and EMPG~10, whose O/H could be explained with model M1 at an age of about 100~Myr. 
We thus computed a model, M2, with a top heavier IMF, with the same chemical evolution parameters as M1 but an IMF slope in the high mass domain $\xUP = 0.6$, while keeping the same   \MUP \ = 100 \Msun. Model M2 is depicted in cyan in Fig.~\ref{pisn1}.
As expected, M2 shows a faster metal enrichment than M1, being able to reach the observed O/H value of EMPG~6 in less than 100~Myr.
However, increasing the fraction of massive stars with a  flatter IMF,  results in a larger production of $\alpha$-elements, with respect to Fe, and the Fe/O ratio drops significantly instead of increasing. Model M2 runs below  model M1  and is not able to reproduce the observed Fe/O ratios of EMPG~3, ~10 and ~6. It is however able to reproduce  the low Fe/O values of the other EMPGs, at an age that is still compatible with that estimated from spectro-photometric and nebular properties of these galaxies.\\ 


As discussed earlier, one possibility to obtain the observed high Fe/O ratios could be by means of selective outflows, in
the specific case a more efficient ejection for O than for Fe. However as long as elements are produced by the same stars (massive)  and on similar (short) timescales, like in the case of the young EMPGs we are discussing, then  they should have comparable ejection efficiencies \citep{recchietal2004A&A}. 
Furthermore, it has also been shown with both simple analytic and numerical models that, if the yields do not depend on the metallicity, the models evolve at constant abundance ratios also if differential (i.e. with efficiency independent from the element) winds are assumed \citep{recchietal2008A&A}.
Thus if the  observed abundance ratios are those produced during the ongoing starburst phase, the high observed Fe/O ratios and the normal observed N/O ratios of EMPG~3 and EMPG~10 must be a consequence of the initial yield ratios \citep[e.g. eq. 2 and 29 in][]{recchietal2008A&A}.

In this respect Fig.~\ref{HW2002} shows that the ejecta from PISN, once integrated on a suitable IMF, may indeed be able to reproduce the observed abundance ratios.
Therefore we now consider models that include also PISN ejecta and, for this purpose we need to extend \MUP \ in the domain of very massive stars.\\ 

By adopting the ejecta from \citet{Goswami2020}, the best model we can obtain with reasonable IMF values is model M3. It is characterized by a top heavy IMF with  $\xUP = 0.6$ and \MUP =300 \Msun. Such a flat slope is close to the lowest value determined for the Arches star cluster \citep{marks_imf_2012MNRAS} or for NGC~2070 in the 30 Dor star forming region \citep{Schneider2018}.
The Fe/O ratio predicted by model M3 is significantly higher than that obtained by all previous models. However, although it is fully compatible with the Fe/O ratio of EMPG~3, considering the large error bar for this object, model M3 falls too short to explain the large Fe/O observed for EMPG 10.
We note also that model M3 is able to reach the (O/H) values of EMPG~3 and EMPG~10 in less than 30~ Myr and that of EMPG~6 in less than 60~ Myr. 
For what said before, the young age obtained, on one side justifies the omission of selective outflows and, on the other, renders the model quite independent from the assumed SNIa fraction.\\

Another evident characteristic of model M3 is the sharp drop of the Fe/O ratio when it reaches 12+log(O/H)$\sim$8. This cannot be a result of the intervening type~Ia supernovae above about 60~Myr because in that case the Fe/O ratio should have increased even further. Instead, it is a clear evidence that the adopted ejecta are not metal independent. Indeed  the \citet{Goswami2020} ejecta were computed by interpolating those by \citet{Heger_Woosley2002y} as a function of \Mhe\, and it is the relation between \Mhe\ and  \Mi\ 
that strongly depends on metallicity because of the adopted mass-loss rates. The general trend is that at increasing metallicity and keeping \MUP\ fixed, the upper IMF spans a decreasing range of \Mhe\ values until \Mhe\ becomes lower than the threshold to produce PISN. Given the dependence of the Fe and O ejecta from \Mhe\, the model shows such a sharp drop in the Fe/O ratio.
Thus, model M3 shows that a continuous burst of star formation actually evolves naturally toward a decreasing sequence, that takes place when the metallicity reaches the {\it stellar evolution} threshold for the PISNe production, i.e. that produced by large mass-loss rates. Since all the selected galaxies have high specific star formation rates and low total stellar masses and estimated young ages, it could also be that some of them form an ageing sequence of the starburst.

\subsection{The N/O ratio}

We have then compared with data the N/O ratios as given by the same models M2 and M3 discussed for the Fe/O ratio. Model M2 performs even worse than model M1. Due to the absence of rotation in the yields, the primary N production is not well reproduced, similar to M1. Moreover, due to the flatter IMF slope, N/O is lower than in M1, as O is produced more as explained above, and the model is only able to follow the secondary behaviour of N at higher metallicities.
Given that N is not a product of the explosive phase and that the ejecta are those of non rotating stars, model M3 behaves similar to M2 but produces even lower values of N/O ratios because PISN enhances the production of O without affecting that of N.

None of the models we have tested so far is able to produce the observed N/O ratio at early times, i.e. at low metallicity.

It has recently been convincingly  shown 
that only massive stars models with rotation are able to produce and eject, even during their pre-supernova evolution, enough N to reproduce the N abundance pattern observed in metal poor stars of the MW and in nearby galaxies \citep{Limongi_etal18, Prantzos_18, Grisoni2021}.

In fact, it has been shown that rotation is needed to reproduce the abundance patterns of other chemical elements in the MW besides nitrogen such as carbon \cite{Romano2020} and fluorine \cite{Grisoni2020b}. Since this is also true for stars that end their life
as PISN stars \citep{,Takahashi2018ApJ857} and since these latter stars seems to be needed to reproduce the high observed Fe/O ratio of EMPG 3,10 and 6, we have computed new VMO models with rotation. 

We remind that, while the final O and Fe ejecta from PISN models are not significantly affected by the initial rotational velocity of the stars, the results are very different for N14, as shown in Figure  \ref{HW2002}. We have thus calculated several sets of VMO models at increasing rotational velocity, from 100 \Msun \ to 350 \Msun \ for Z$_i = 0.0001$ and Z$_i=0.001$. 
The stellar models have been followed from the pre-main sequence phase until the beginning of central Oxygen burning when they become unstable due to efficient electron-positron pair production.
At this stage N production by these stars ceases because the evolutionary timescales become very short and wind ejecta inefficient. The total ejecta are then obtained by adding the ejecta of explosive pure He models \cite{Heger_Woosley2002y}, matched by means of the PARSEC final He core mass.
Starting from a negligible N production in non rotating models,  the N ejecta increases at increasing rotation, becoming significantly higher when the initial angular rotational velocity, $\Omega$, reaches 60\% of the  critical  one, $\Omega_{crit}=(2/3)^{3/2}\sqrt{GM/R^3}$, the latter corresponding to the point where the centrifugal acceleration equals the gravitational one at the equator.\\



\begin{table*}[t]
\footnotesize
\caption{Chemical evolution model parameters and properties of the initial VMO burst when the gas metallicity reaches Z=\Zcr.}
\label{tab_vmozcr}
\centering
\tiny
\begin{tabular}{|c|c|c|c|c|c|c|c|c|}
\hline
\multicolumn{3}{|c}{ \Mchar = 200 \Msun } &  
\multicolumn{2}{c}{${k}$=1} &
\multicolumn{2}{c}{${\tau_{\rm inf}=0.1}$ } &
\multicolumn{2}{c|}{${A_{\rm SNIa}}$=0.04} \\
\hline
\multicolumn{1}{|c|}{{Name}} &
\multicolumn{1}{|c|}{$\nu$}& 
\multicolumn{1}{|c|}{{\Zcr}} &
\multicolumn{1}{|c|}{t$_{\Zcr}$} &
\multicolumn{1}{|c|}{{\Ohl}} &
\multicolumn{1}{|c|}{{M$cum$}} &
\multicolumn{3}{|c|}{{N$_{VMO}$} }\\
\hline
& Gyr$^{-1}$ & Z/\Zsun & Myr&  & $M_\odot$&Model &
EMPG3$^a$& EMPG10$^a$ \\
\hline
\Mqot&1 &1.0E-4 & 2.77 &4.520& 3.814E4 &   176 & 194&	393 \\
\Mqoq&1  &1.0E-4 & 2.78 &4.580& 3.834E4 &   177 & 194&	394 \\
\Mcot&1  &1.0E-2 & 5.01 &6.805& 1.232E5 &   569 & 720&	1784 \\ 
\Mcoq&1  &1.0E-2 & 5.13 &6.815& 1.292E5 &   597 & 741 &	1807 \\ 
\Msot&5  &1.0E-4 & 2.70 &4.692&3.615E5& 1672& 2122& 4526  \\
\Msoq&5  &1.0E-4 &2.70&4.663&3.596E5& 1675& 2100& 4454\\
\hline
\end{tabular}
{\raggedright \par $^a$Computed from the model normalized to the observed stellar mass at the corresponding value of \Ohl. \par }
\end{table*}

\section{Chemical evolution models with a variable IMF}
\label{cheIMF}

Besides the problem of the N abundance, even model M3, which includes the PISN yields, is not able to fully reproduce the high Fe/O ratios observed in EMPG~10. 
It also has a prolonged phase of relatively high Fe/O until a metallicity of 12+log(O/H) $\sim$ 8 is reached. Beyond this point, higher mass-loss rates caused by the high metallicity prevents stars with high He core mass to enter the domain of PISN  and Fe is no more produced while O still continues to be copiously ejected.
To sample the effects of VMO exploding as PISN we include in our chemical evolution model an early phase where the IMF has the following form: 
\begin{equation}\label{eqwise}
\phi(\Mi) =  \frac{dn}{d\log(\Mi)} \propto \Mi^{-1.3}\times~exp\left[\,- \left(\, \frac{Mchar}{Mi} \right)^{1.6}\right]\,  
\end{equation}
with  $80 \Msun \leq  Mi \leq 350$ \Msun .

This distribution, originally devised by \citep{chabrier2003}, has been adopted by
\cite{wiseetal03} to describe the masses of population III stars, with a characteristic mass \Mchar = 40  \Msun. Here we test a few values of \Mchar\  between   \Mchar = 200  \Msun\ and \Mchar = 300 \Msun,  because the relative PISN ejecta of O and Fe strongly depend on their He core mass in the above range of initial masses (see Fig. \ref{HW2002}).\\
This phase of PISN enrichment is confined to the early evolution of the starburst, defined by the gas metallicity being below a threshold metallicity \Zcr. Above \Zcr\ we adopt a Kroupa-like IMF, Eq.~\ref{imfkroupa}, with \xUP=1.3  \ and \MUP=40  \Msun.\\
With this bi-modal IMF we test two different values of the threshold metallicity, \Zcr\ $=$ 10$^{-4}$ and 10$^{-2}$ \Zsun. As discussed in the Introduction, these values have been suggested by \citet{Schneider2006MNRAS} and \citet{salvadorieial2008MNRAS}, respectively for the cases of efficient and inefficient dust cooling during the early star formation process.

The results are presented for yields obtained adopting stellar evolution models with two different rotational parameters, $\omega$=$\Omega/\Omega_{crit}$=0.3 and 0.4 (because with these values we bracket the observed N/O ratios), 
and for \Zcr  = 10$^{-4}$ (models \Mqot \  and \Mqoq \ ) and \Zcr  = 10$^{-2}$ (models \Mcot \  and \Mcoq.). 
The other parameters of the chemical evolution models for the two \Zcr \ values are reported in Table \ref{tab_vmozcr}.

\begin{figure} \centering
\resizebox{0.95\hsize}{!}{\includegraphics[angle=0]{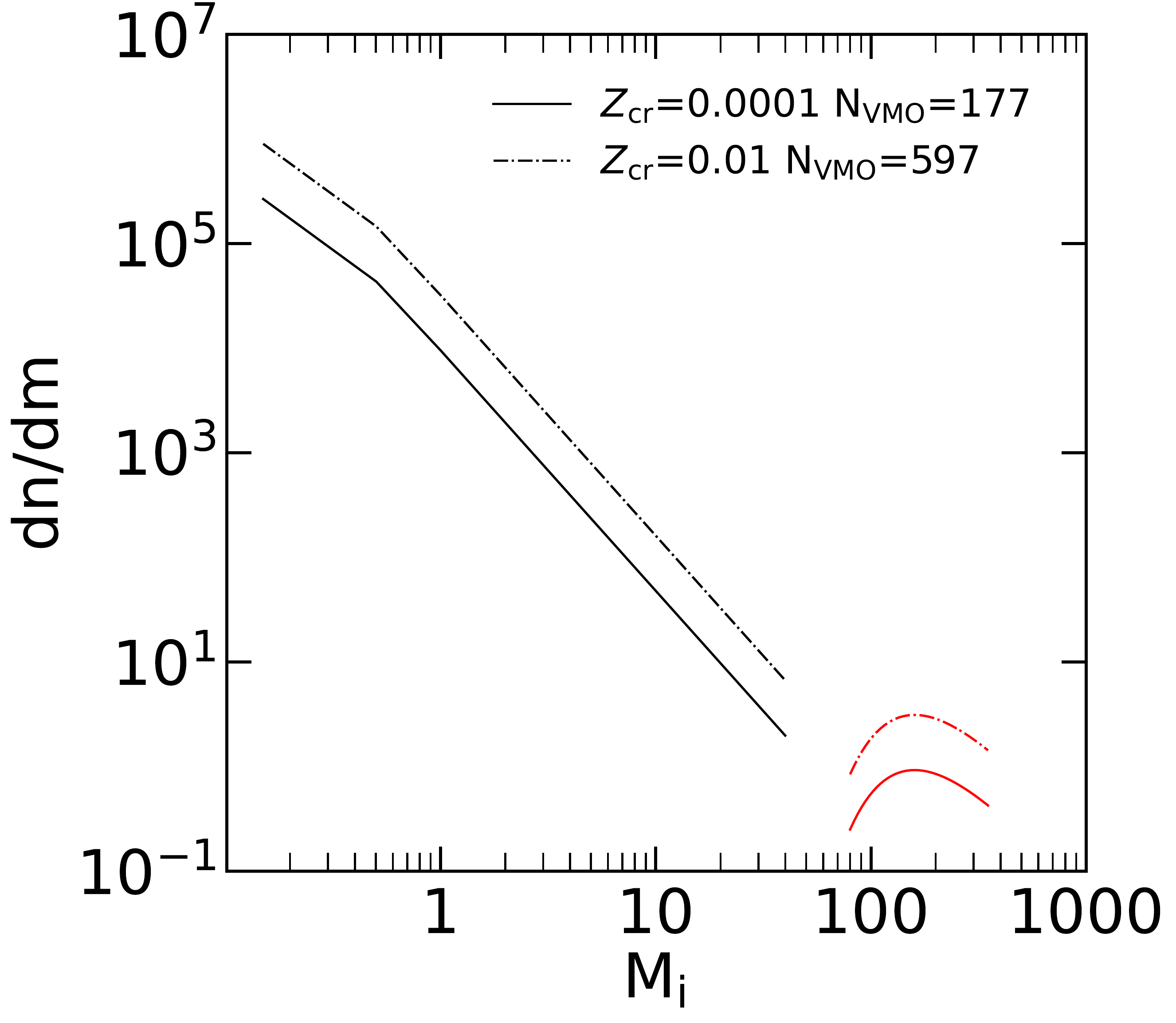}}
\caption{
The adopted IMF for the chemical evolution of models \Mqoq \  (solid red line) and \Mcoq \ (dot-dashed red line). The VMO IMF distributions are given by Eq.~\ref{eqwise} with \Mchar = 200 \Msun \ and normalized to the cumulative mass reached at the corresponding \Zcr.
For comparison, the IMFs of the post-VMO bursts phases, with the same mass normalization, are also shown (black lines). The expected number of VMOs is shown in the label and in Table~\ref{tab_vmozcr}.}
\label{zmchar200imf} 
\end{figure}

The adopted bi-modal IMF is shown in Fig.~\ref{zmchar200imf}. The VMO IMF, from 80 \Msun to 350 \Msun \ is normalized to the integral of the SFR from the beginning to the time when the total metallicity reaches the value Z= \Zcr $\times$ \Zsun \ for \Zcr = \ 0.0001 (solid line) and   \Zcr = \ 0.01 (dashed line), respectively. This corresponds to the total mass of gas, converted into stars, \Mcum \ , by the time \Zcr  \ is reached. The value of \Mcum \ at Z/\Zsun =\Zcr \ is shown in Table \ref{tab_vmozcr} for the different chemical evolution models.\\
The total number of VMO 
stars that have been formed, \Nvmo, are indicated in the figures and reported in the last column of Table \ref{tab_vmozcr}. After \Zcr\ is reached, the IMF is changed to the Kroupa-like IMF. This is also plotted in Fig.~\ref{zmchar200imf}, normalized to the corresponding value of  \Mcum \  reached at \Zcr , for sake of comparison.

The last two columns, in Table \ref{tab_vmozcr},  refer to the number of VMOs at Z/\Zsun =\Zcr\ expected from a model whose total stellar mass equals the  observed galaxy stellar mass, at the corresponding value of \Ohl.\\
\begin{figure} 
\centering
\resizebox{0.95\hsize}{!}{\includegraphics[angle=0]{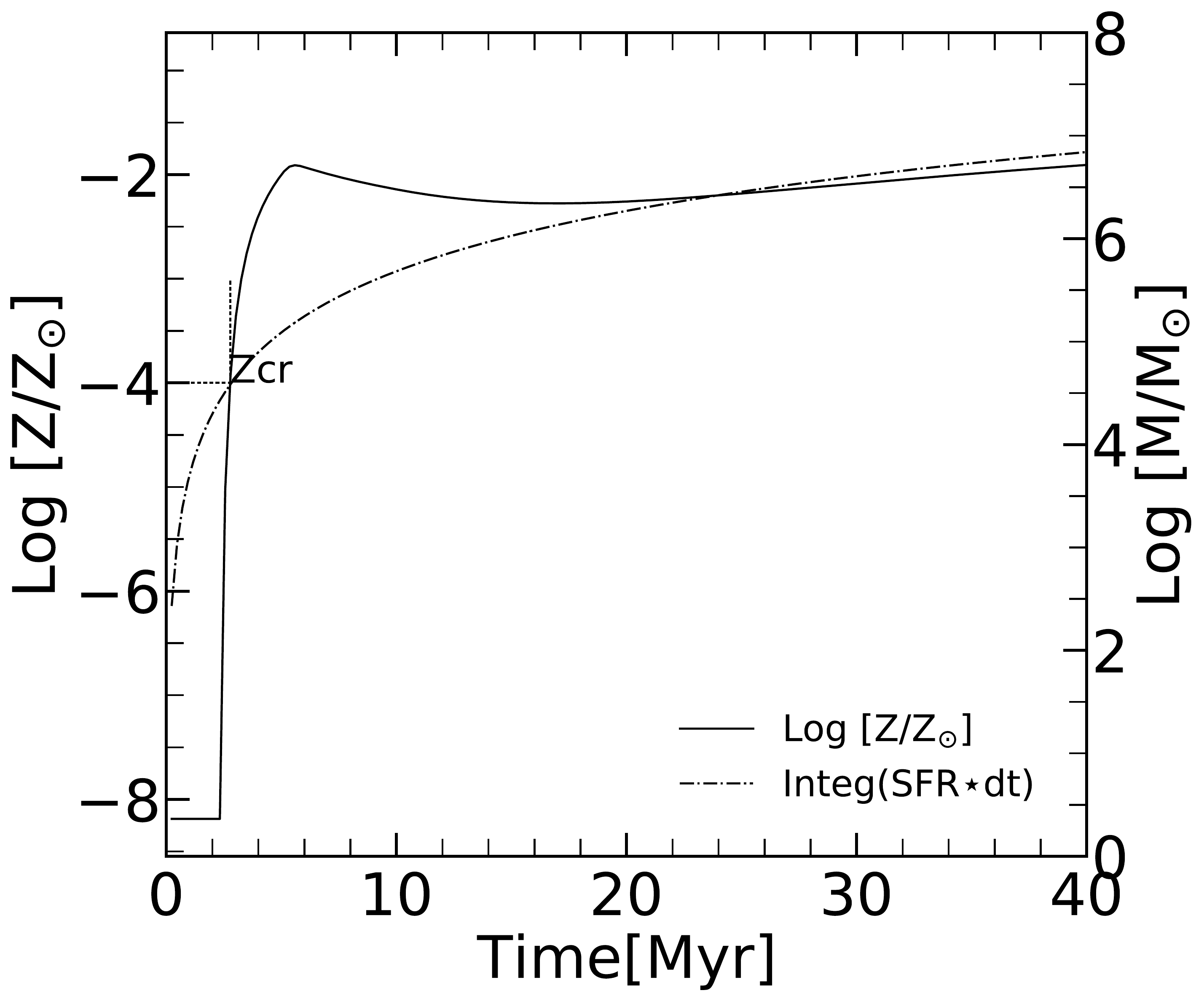}}\\
\resizebox{0.95\hsize}{!}{\includegraphics[angle=0]{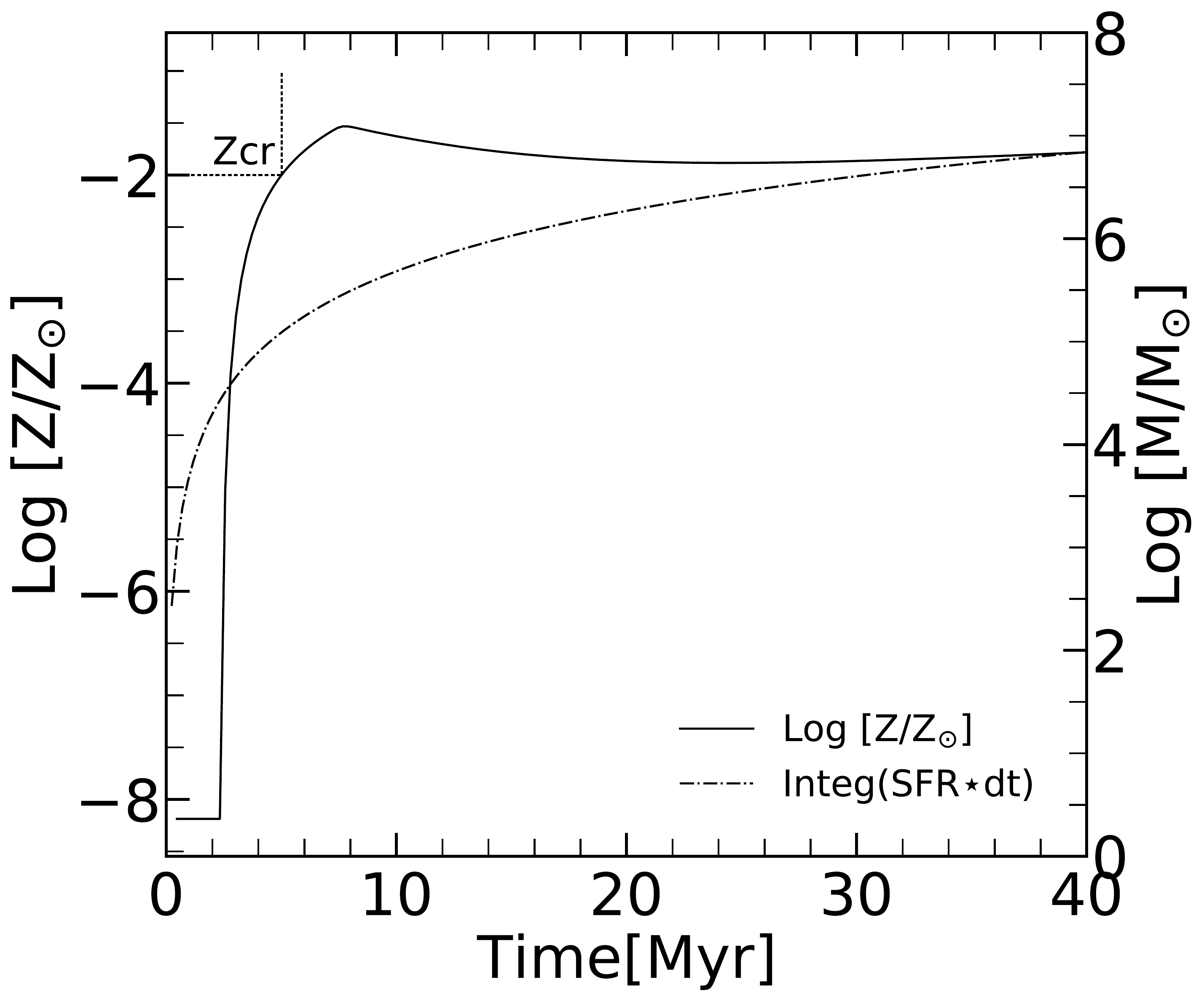}}
\caption{Evolution of the gas metallicity (solid lines) and of the cumulative mass converted into stars (dash-dotted lines) of models  that include an early burst of VMOs exploding as PISN. The threshold metallicity for VMO formation is Z$_{cr}$ = 10$^{-4}$ (\Mqoq, upper  panel) and 10$^{-2}$ (\Mcoq, lower panel). Their IMF distribution is shown in Fig.~\ref{zmchar200imf}. In both panels the adopted \Mchar\ is 200 \Msun. See text for details.}
\label{zmchar200} 
\end{figure} 


\begin{figure*} 
\centering
\resizebox{0.41\hsize}{!}{\includegraphics[angle=0]{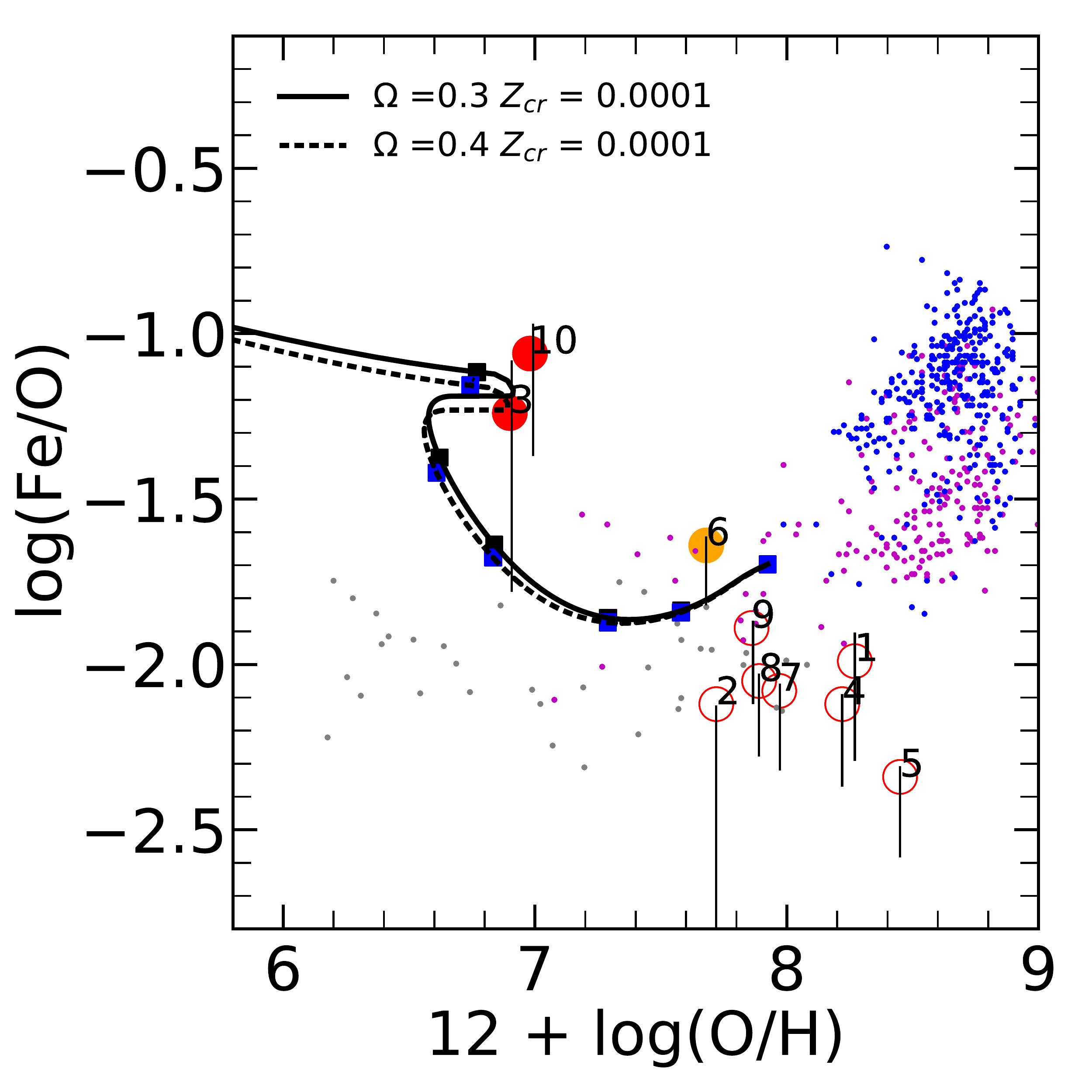}}  \   \
\resizebox{0.41\hsize}{!}{\includegraphics[angle=0]{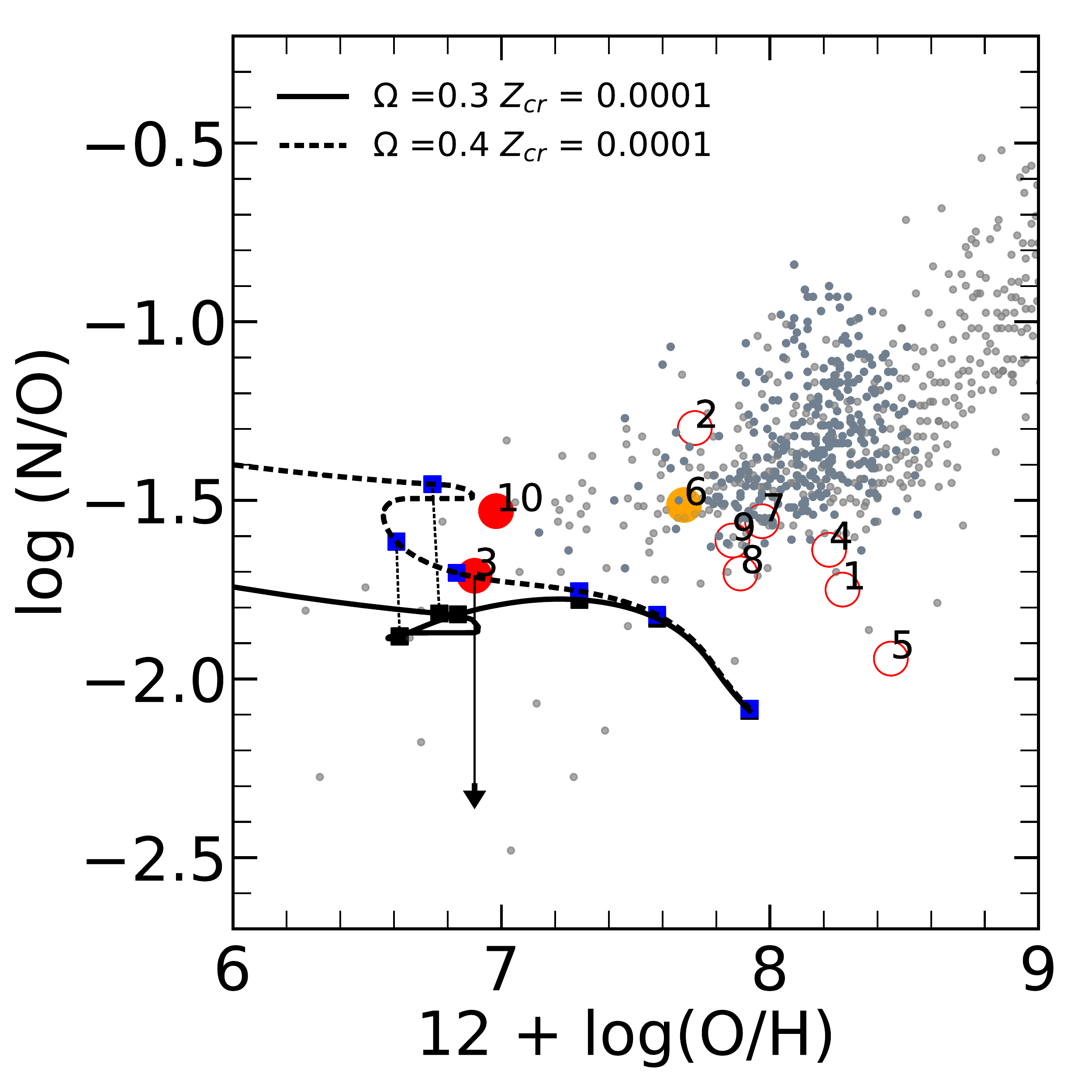}}\\
\resizebox{0.85\hsize}{!}{\includegraphics[angle=0]{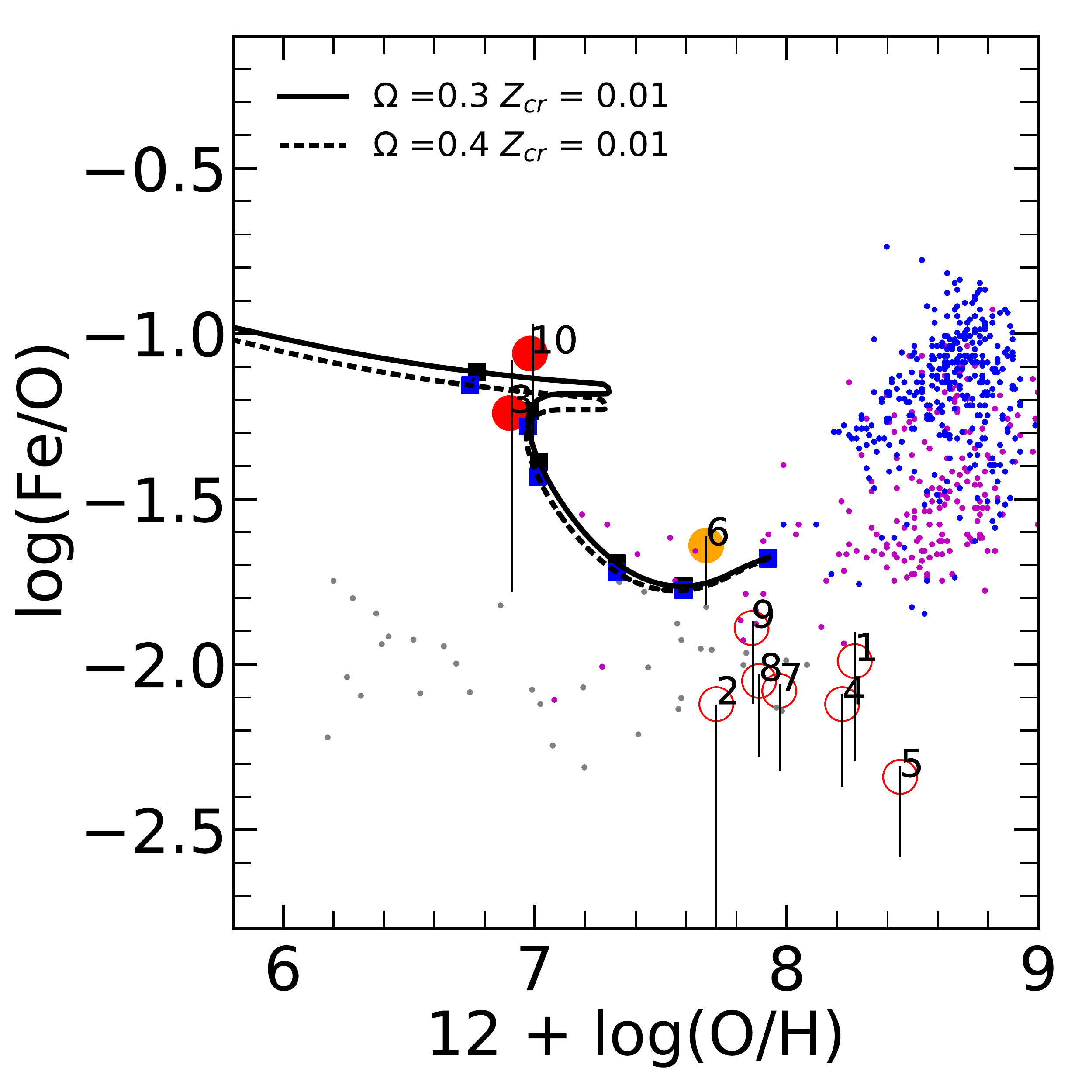} 
\includegraphics[angle=0]{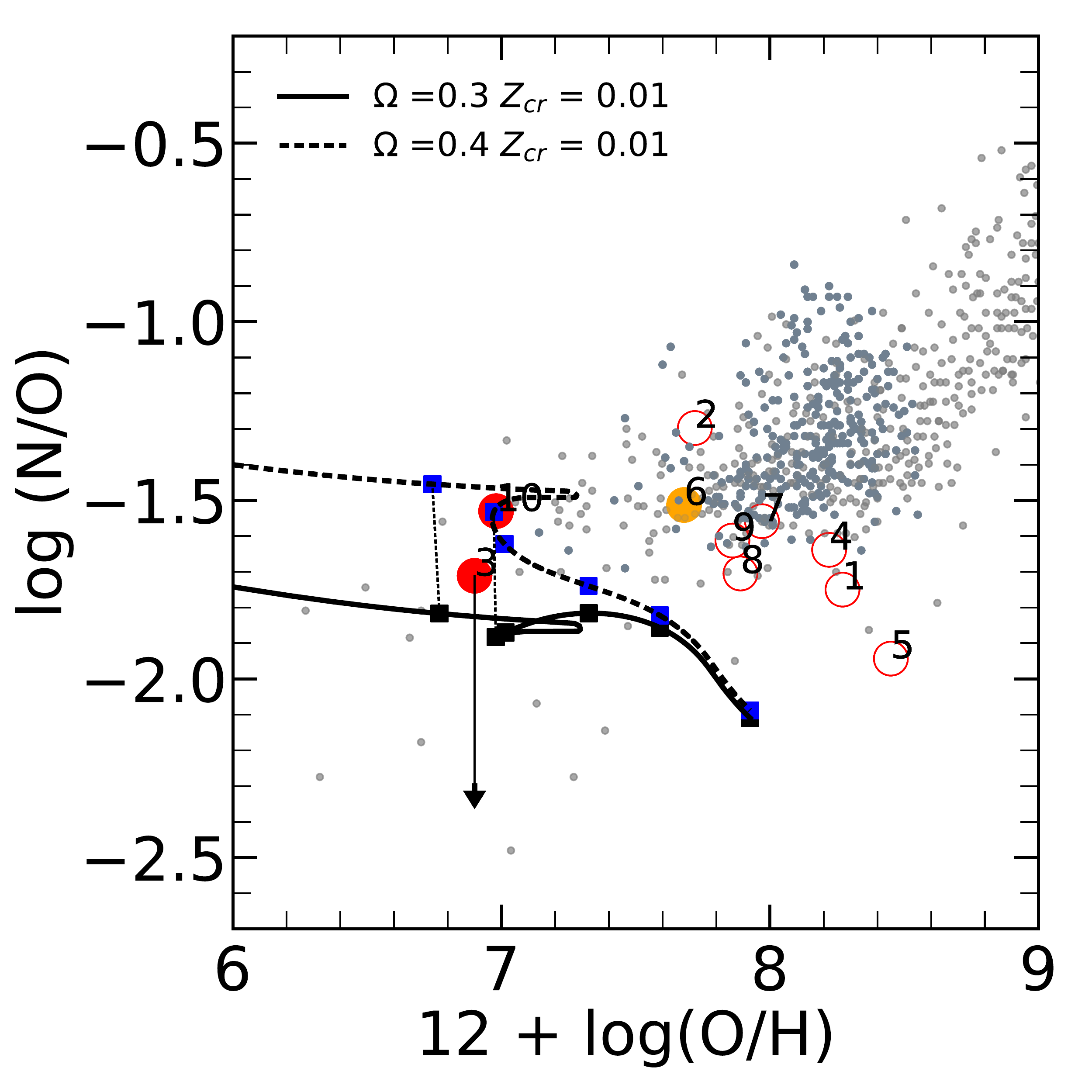}}\\
\caption{Effects of including rotational yields from PISN in the early evolution of the Fe/O (left panels) and N/O (right panels) as a function of the metallicity 12+log(O/H).
The upper panels refer to Z$_{cr}$ = 0.0001, the lower panels refer to Z$_{cr}$ = 0.01. Symbols are the same as in Fig. \ref{pisn1}.} 
\label{femchar200} 
\end{figure*}

Figure \ref{zmchar200} shows the early evolution of the metallicity of the starburst model \Mqoq , for  \Zcr = 0.0001 (upper panel), and \Mcoq ,  and \Zcr = 0.01 (lower panel), both for \Mchar = 200 \Msun.
They are computed with the same yields obtained with PISN models with rotation parameter  $\omega$=0.4. The models \Mqot \ and \Mcot , computed with lower rotational velocity,
$\omega$=0.3, run very similar to the previous ones and are not shown in the plot for sake of simplicity.\\
All the new models  begin from a metal free gas, Z=10$^{-10}$.
The metallicity (solid lines) remains flat until the first PISN explode, after about 2.3~Myr.
This is the typical lifetime of a VMO, even  at zero metallicity. After this age, with the adopted chemical evolution parameters, the metallicity rises very fast. Model \Mqoq\   reaches  \Zcr = 0.0001 almost immediately, challenging a further formation of VMOs. However, during the first 2.3 Myr, already about 180 VMO stars have been formed, out of a cumulative stellar mass formed (dot-dashed line) of about  \Mcum =  38000 \Msun . The models metallicity reaches \Zcr\ at an age of t$_{\Zcr}\sim$2.8~Myr. At this point, the IMF turns to a Kroupa-like IMF beginning to form stars with \Mi $\leq$ 40 \Msun. However, the  metallicity enrichment from this VMO burst is effective until an age of  t$_{\Zcr}$ + 2.3 Myr is reached. Thus the metal enrichment continues up to about 5~Myr and after this age the metal enrichment is only due to massive stars with \Mi $\leq$ 40 \Msun.
The value of the O abundance reached at \Zcr =0.0001, shown in Column 4 of Table \ref{tab_vmozcr}, is \Ohl $\sim$ 4.5, i.e.\ about two orders of magnitude less than that observed in EMPG 3 and EMPG 10. However, the further enrichment by the youngest VMOs pushes the value of  \Ohl\ up to that observed in EMPG 3 and EMPG 4 at an age of about 5~Myr. After this age, the metallicity shows a slight decrease until the enrichment by less massive stars begins to compensate the dilution due to infalling gas. During this phase \Ohl\ decreases by 0.4~dex.\\
The evolution of Model \Mcoq\ is very similar to that of Model \Mqoq, though with different timescales due to the larger  \Zcr\ =0.01 adopted. Model \Mcoq\ reaches \Zcr\ at an age of t$_{\Zcr}\sim$5~Myr with  \Ohl $\sim$ 6.8. This value is almost the observed one in EMPG 3  and EMPG 10. This could indicate that, in this scenario, EMPG 3 and EMPG 10 just reached the critical threshold for PISN production. In  this model, \Mcum\ and  \Nvmo\ at \Zcr\  are 1.35 $\times$ 10$^5$ \Msun\ and 600, respectively. Even in this model the metal enrichment by VMOs  lasts for other 2.3~Myr and then the metallicity begins to decrease, showing a  \Ohl\ decreases by 0.4~dex.
\\

The evolution of the Fe/O and N/O ratios as a function of \Ohl\  of the models including PISN yields, is shown in Figure~\ref{femchar200}. The observed data are the same as in Fig.~\ref{pisn1}. 
The starburst ages derived by \cite{Kojima2021} from spectral synthesis modelling are young for all the galaxies plotted in Figure~\ref{femchar200}, with ages $\leq$ 100 Myr. This implies that the metallicity in these objects, e.g. \Ohl, must grow very quickly, possibly requiring a top heavy IMF. Thus, comparing the left panels of Figs.~\ref{femchar200} and \ref{pisn1} we clearly see that an early burst of VMOs is needed to reproduce the observed high values of the Fe/O ratio and of the  \Ohl, at early times. For the N/O ratio, these conditions are not even enough. Looking to the right panels of the figure, we see that rotation is indeed an essential ingredient to reproduce Nitrogen enrichment at very low metallicity.\\ 

We note that, while Fe yields are only mildly affected by rotation, N yields do strongly depend on that \citep{Limongi_etal18}. 
We remind that this conclusion for the Fe yields by PISNs is mainly based on the results obtained by \cite{Takahashi2018ApJ857} who computed both the pre-SN evolution and the following explosive models for models with and without rotation (e.g. Fig.~\ref{HW2002}). For N we know that only the pre-SN evolution conditions matter and that other parameters beside rotation may affect its yields, in particular that describing the envelope overshooting \citep{Costaetal2020}. 
In order to reproduce the observed N/O ratios of EMPG 3 and EMPG 10, we need to use massive star models with an initial rotation parameter between $\omega$=0.3 (solid lines) and $\omega$=0.4 (dashed lines). \\
In all the plots, the solid squares along the models mark ages of 5, 20, 30, 60, 100, 200 Myr, with the latter being the rightmost point. 
In the Fe/O diagrams, EMPG 3 and EMPG 10 can be well fitted  with both the cases of \Zcr=0.0001  and \Zcr=0.01 and for any rotation value. The time scale of the enrichment indicates, at least for EMPG 10, an age slightly larger than 5~Myr, which is compatible with the very young ages derived from spectrophotometry  \citep{Kojima2021}.
EMPG 3 could also be compatible even with  a lower Fe/O ratio but, in this case the age would be older but still in the observed range.
 
In the case of \Zcr\ = 0.0001, the N/O ratio of EMPG 10 is marginally fitted with the higher rotation young model. \\
For EMPG 3 we have only an upper limit on the N/O ratio and model \Mqoq\  fits its position very well in both diagrams. However, this model requires two quite different ages for Fe/O and N/O, respectively. The best solution  for \Zcr\ = 0.0001 is provided by model \Mqot\ with an age $\geq$  5~Myr. \\
By relaxing the critical metallicity to \Zcr\ = 0.01,  more solutions are possible. EMPG~10 can be well fitted by the \Mcoq\ model at an age $\geq$ 5~Myr or even at an age $=$20~Myr. While for EMPG~3 the best model is the \Mcot\ again for an age  $\geq$  5~Myr 
or even at an age of $=$20~Myr. \\
In any case, the presence of an initial burst of VMOs can explain the data, as assumed in models \Mqot, \Mqoq, \Mcot\ and \Mcoq.
These models show also another interesting feature, which is their rapid Fe/O decline at increasing  \Ohl , as a consequence of the suppression of the VMO burst, when the critical metallicity is reached.
This feature is visible also in model M3 (Fig. \ref{pisn1}), but in this case, as discussed above, the threshold metallicity, \Zcrse\, for the suppression of the PISN channel depends on stellar evolution, when large mass-loss rates are boosted by a high metallicity. In our models and for an \MUP \ =350 \Msun , \Zcrse\ $\sim$0.4 \citep{Goswami2020}. \\
The metal poor galaxies analysed by \cite{Kojima2021} actually show a trend of declining Fe/O at increasing \Ohl , as shown in Figure~\ref{femchar200}. However this trend cannot be explained by the models presented in that figure if, together with the abundances, we consider that the observed galaxies are all dominated by active starbursts with ages below 50~Myr.
The oldest of our model has already an age of 200~Myr and marginally reaches the bulk of the remaining metal poor galaxies in the Fe/O diagram while it cannot reach them in the N/O diagram.
We note however that many of them show Fe/O and N/O ratios that are even below those observed among the bulk of low metallicity stars. While low values of N/O are expected among non-rotating models, also the low Fe/O ratios could challenge standard chemical evolution models, as it may be seen in Figure~\ref{pisn1}. \\
However, the relatively low Fe/O ratio  of "evolved" young metal poor starburst galaxies could be explained by a combination of an early burst of VMO, like for EMPG 3 and EMPG 10, and a population of stars that produces the oxygen rich/iron poor ejecta. This could have the twofold effect of favoring a fast \Ohl\ evolution and, at the same time, a rapid decline of the Fe/O ratio. Such a model can be  obtained by combining an early burst of VMO stars, like in the case of models \Mqot, \Mqoq, \Mcot, \Mcoq, and a model that includes a top-heavy IMF with \MUP\ $\leq$  200 \Msun.
For the VMOs component we adopt the same form of the \Mqot\ IMF and, when \Zcr\ $=$ 0.0001 is reached and the IMF turns into a standard Kroupa-like one, the formation of a small fraction low mass VMOs (M$_i \leq$ 150 \Msun ) is still allowed, with an exponent \xUP\ $= 0.9$. Furthermore, in order to comply  with the fast enrichment required by the age estimates performed by \cite{Kojima2021}, we need to use a relatively high  star formation rate efficiency $\nu = 5$. The properties of the considered models, \Msot\ and \Msoq, are in Table \ref{tab_vmozcr}. 

\begin{figure*} 
\centering
\resizebox{0.41\hsize}{!}{\includegraphics[angle=0]{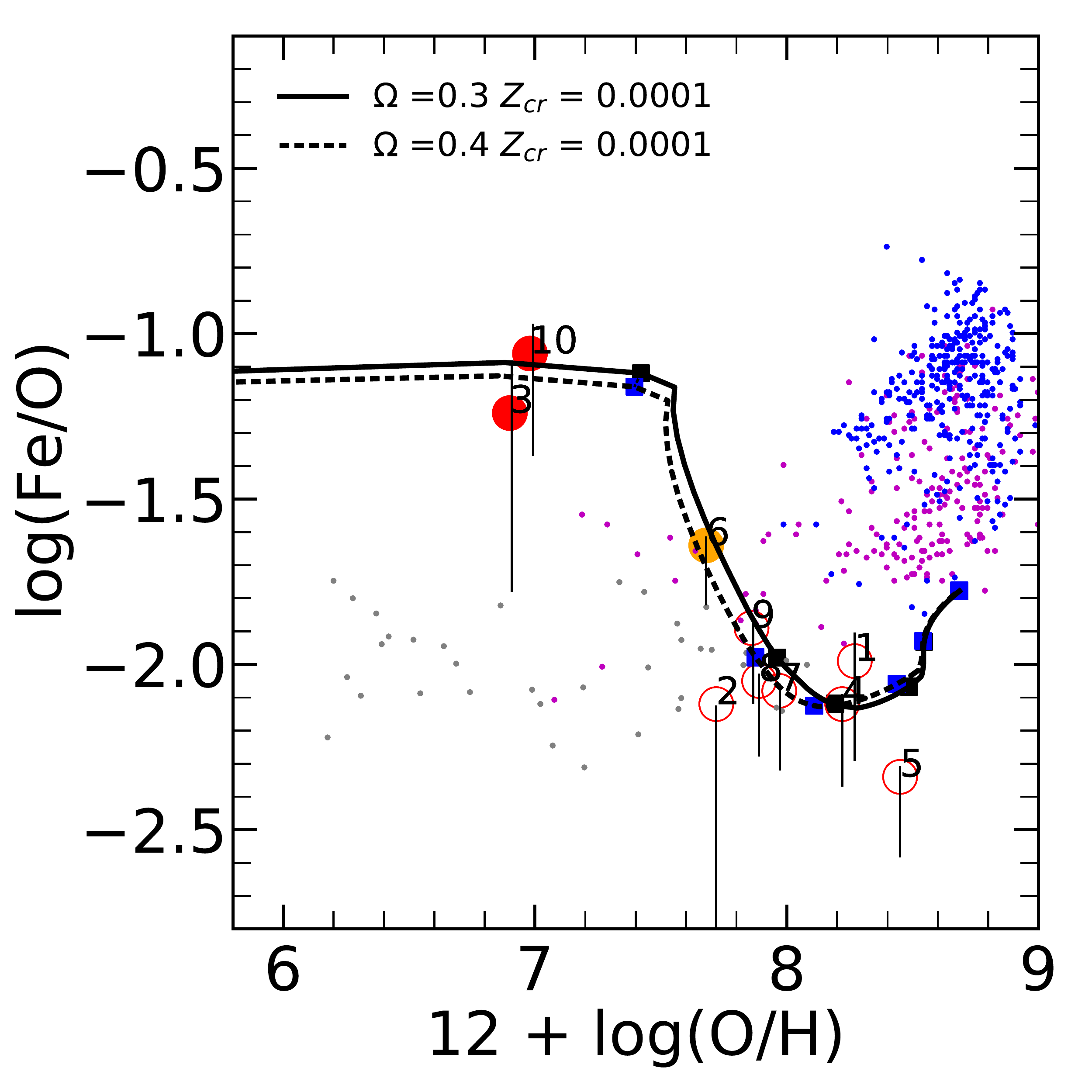}}  \   \
\resizebox{0.41\hsize}{!}{\includegraphics[angle=0]{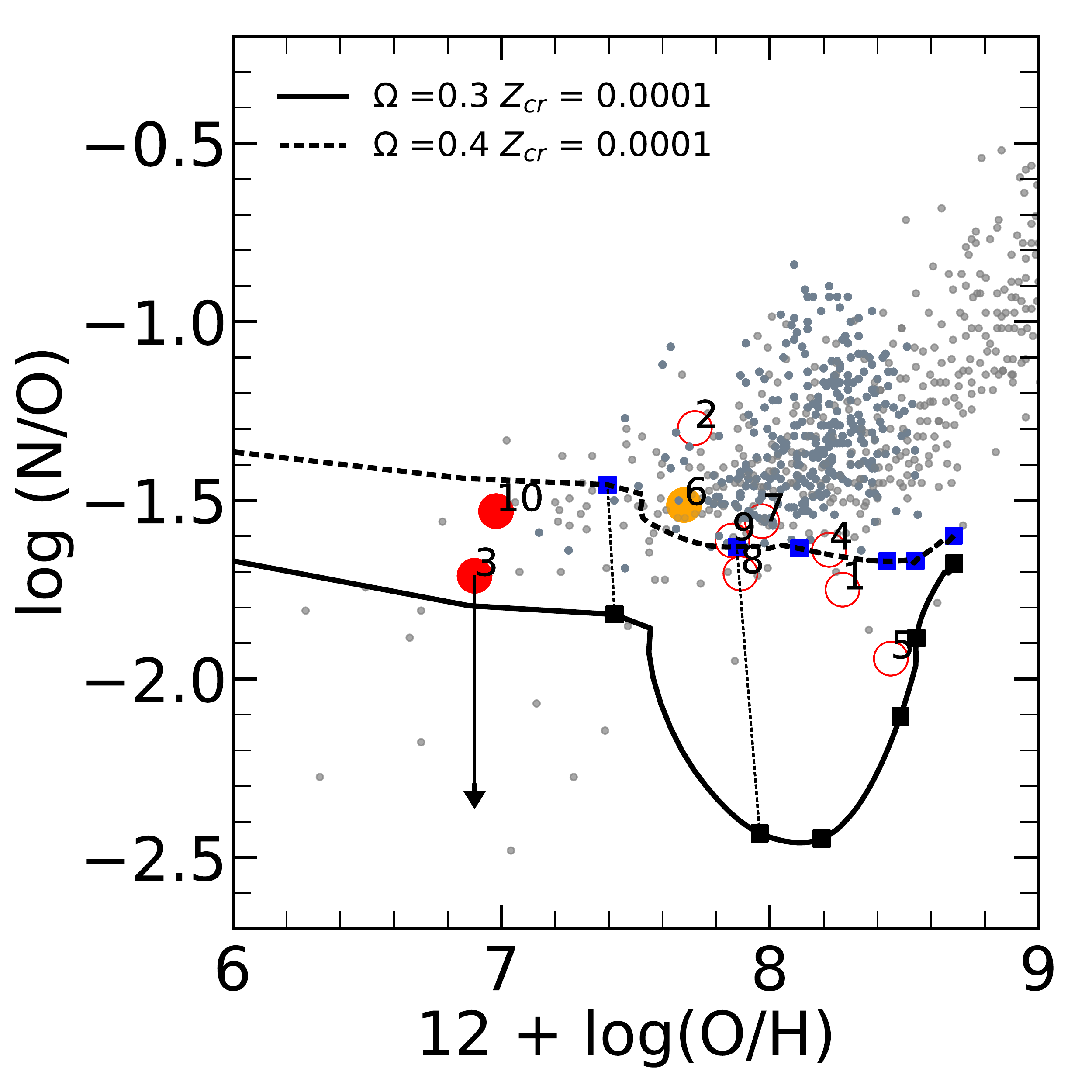}}\\
\caption{Fe/O (left panel) and N/O (right panel) in models \Msot \ (solid lines) and \Msoq \ (dashed lines) as a function of the metallicity 12+log(O/H).
The panels refer to \Zcr\ = 0.0001 and the symbols are the same of Fig. \ref{pisn1}.} 
\label{mpgfemchar200} 
\end{figure*}

The evolution of the Fe/O and N/O ratios obtained with these  models is shown in Figure~\ref{mpgfemchar200}. Their metallicity \Ohl\ rises rapidly up to the value corresponding to \Zcr\ = 0.0001. Beyond this point, the Fe/O ratios, that initially correspond to those of EMPG 10, begin a rapid decrease, reaching the value observed in EMPG 6. Later, at an age between 20 and 60 Myr, both models reach the position occupied by the other EMPRESS galaxies.
Only EMPG 5 cannot be reproduced by these models.
However we remind that, at the higher metallicities, the estimate of the Fe/O ratios is generally more uncertain because of selective dust depletion \citep{Rodri2005,Izotov2006}. 
As far as the N/O ratio is concerned, only model \Msoq\ is able to fit the observations while model \Msot\ has a too low N/O ratio.


\section{Conclusions}\label{conclu}

In this work, we analysed the recent observations 
of EMPGs galaxies with peculiarly high Fe/O and young starburst ages \citep{Kojima2021}.
We initially investigated these peculiarities using the recent yields from \cite{Goswami2020}, which  take into account the later evolutionary stages of non rotating massive stars, such as PISN and PPISN. We run different chemical evolution models with different IMF upper mass limits and slopes. 
We have shown that in order to reproduce the high  Fe/O ratios observed at 12+log(O/H) $\sim 7$ in EMPGs 3 and, in particular, in EMPG 10, we need to include the effects of PISN, as in model M3. Other solutions are excluded because the early evolution of such starbursts keeps information of the ejecta of the most massive stars, and only PISN coming from a VMO distribution
that produces the right He core masses is able to produce the right ejecta.
However, we also show that, in order to reproduce the observed primary N behaviour of these EMPGs, we need to consider stars with suitable rotation. Indeed model M3, with
a top heavy IMF up to 300 \Msun\ and \xUP $=0.6$, while reaching a large Fe/O ratio, on one side it is not able to reach that of EMPG 10 and, on the other, it shows a too low N/O ratio that decreases even more at increasing contribution of non rotating PISNs. 

We thus complemented \cite{Goswami2020} yields with new very massive star models with  different initial rotational velocities, from $\omega$ =0.1  to $\omega$=0.6. In rotating massive stars nitrogen is a product of pre-supernova evolution that is barely changed by the final explosion. Our calculations show that N production from the PISN depends very strongly on the initial degree of rotation. While without rotation the N contribution is negligible, it becomes stronger at increasing rotational velocities \cite[see also][]{Takahashi2018ApJ857}.\\
Furthermore, to properly account for PISN in the chemical evolution code, we adopted a bi-modal IMF. During the early evolution, the IMF is skewed towards VMOs  following the prescriptions of \citet[][see Eq. \ref{eqwise}]{wiseetal03}.
Since, according to existing theoretical  investigations, VMOs producing PISN may form only in very metal poor environments, with Z/\Zsun\ $\leq$\Zcr, we assume also that, when the metallicity encompasses \Zcr, the IMF changes to a standard Kroupa IMF (Eq. \ref{imfkroupa}). We have considered two values of the  critical metallicity, \Zcr\ $=$  0.0001 and \Zcr\ $=$  0.01, corresponding to efficient and inefficient dust cooling, respectively \citep{salvadorieial2008MNRAS,Schneider2006MNRAS}.\\
With such a bi-modal IMF and a yield family for different rotational velocities, we searched for suitable chemical evolution models that could explain the main characteristics of EMPG~3 and EMPG~10 namely, (i) relatively high  \Ohl\ at very young ages,  
(ii) high Fe/O ratios and (iii)
 "standard", i.e. typical of low metallicity stars, N/O ratios.
The first point requires a rapid and efficient enrichment. We adopted short star formation and infall timescales, $\nu=1$ Gyr$^{-1}$ and t$_{inf}=0.1$ Gyr, respectively. The second point can be obtained only by means of a suitable population of PISN for which we adopted the IMF of Eq. \ref{eqwise} with \Mchar\ = 200 \Msun. Other values of \Mchar\ between 200 \Msun\ and 300 \Msun\  produce almost the same results while, for \Mchar\ $<$ 200 \Msun, the Fe production is too low. The third point requires models with rotation. We find that, with initial rotation parameters $\omega$=0.3  and  $\omega$=0.4, we could  bracket the observed N/O ratios of EMPG~3 and EMPG~10, as shown in Figure \ref{femchar200}.
Though the threshold metallicity with \Zcr\ = 0.0001  is very low, our models \Mqot\ and \Mqoq\  are able to fit both EMPG galaxies at the observed metallicity, \Ohl, and at young ages.
When \Zcr\ is reached, the number  of VMO formed after normalization to the observed masses of EMPG~3 and EMPG~10, is 194 and 394, respectively. 
Considering the relative evolutionary timescales, these stars should explode with an average rate of one per 20000~yr and one per 7000~yr in EMPG~3 and EMPG~10, respectively.\\ 
A common feature of the EMPRESS galaxies is that they are dominated by
very young starbursts with high specific star formation rates. Contrary to EMPG 3 and EMPG 10, most of them  are characterized by a low Fe/O ratio, which is generally near or even below the lower envelope of the stellar data. Models M1, M2 and  M3 cannot reproduce these peculiarities. This is true also for models with rotation, as can be seen  in Figure \ref{femchar200}. Nevertheless, their low Fe/O could be indicative of an excess of Oxygen, produced by the less massive PISN  as can be inferred from Fig. \ref{HW2002}. We have thus searched if some variants of models M4 and M5 could provide a viable explanation. In order to comply with the young ages observed we performed few tests by increasing the star formation efficiency to $\nu$=5 Gyr$^{-1}$. This is not enough however. The only galaxy that could be explained is EMPG 6 which has a Fe/O value well within the range observed in metal poor stars. Instead, if  we allow   the secondary Kroupa IMF to extend up to \MUP\ =  150 \Msun\ with \xUP\ = 0.9 , i.e. a top heavy Kroupa IMF, we obtain  a model, \Msoq ,  that can match also the positions of the other EMPRESS starbursts at relatively young ages, in both the Fe/O and N/O vs \Ohl \  diagrams, as shown in Fig. \ref{mpgfemchar200}.\\  
We recall that an IMF favouring the formation of massive stars in regimes of bursting star formation has been already claimed in the past \citep{marks_imf_2012MNRAS,jerabkova2018,Zhang2018Natur}. Moreover, we have already shown that there is strong evidence of a well populated IMF up to \MUP \ = 200 \Msun \ in 30 Doradus, with an estimated \xUP \ $\sim$ 0.9  \citep{Crowther2010, Schneider2018}.
In summary, several peculiarities observed among very metal poor starbursts could be explained within a single scenario, in which a high specific SFR produces a fast self-enrichment that drives them first into a phase of high Fe production, and then quickly into the phase of high O production.
Thus, if PISN arise only in high specific SFR objects, this fast evolution could challenge their detection in the local universe 
\\
In this respect, we note that the imprinting left by very massive stars, not only in terms of chemical composition \citep{Takahashi2018ApJ857} but also of chemical evolutionary timescales, could be another important signature to establish the nature of the upper tail of the IMF and its deviations from universality \citep{elmegreenetal_imf_2003,Kroupa2008, marks_imf_2012MNRAS,hoseketal_IMF2019,Romano2020}.   
Finally, we recall that recent gravitational waves detection \citep{Abbottetal2016,Abbottetal2020}
has drawn the attention to a large variety of phenomena that 
could affect the estimate of the initial masses and  metallicities of PISN progenitors and their yields, arising both from current uncertainties in single stellar evolution theory \citep[e.g.][]{Costaetal2020}, possible effects of rotation \citep[e.g.][]{Takahashi2018ApJ857} and also binary interaction \citep[e.g.][]{Han2020RAA,Speraetal2019MNRAS,Stanway_2018,Hurleyetal2002}. 
In summary, a variation of the mass and metallicity domain from where PISN might arise will immediately impact the choice of the IMF parameters needed to reproduce EMPGs observations. 
By converse, the study of these galaxies could have a great impact on our understanding of the uncertainties still affecting stellar evolution models.


\begin{acknowledgements}
We thank the anonymous referee for her/his useful suggestions. This work has been partially supported by PRIN MIUR 2017 prot. 20173ML3WW 001 and 002, `Opening the ALMA window on the cosmic evolution of gas, stars and supermassive black holes'. GC acknowledges financial support from the
European Research Council for the ERC Consolidator grant DEMOBLACK,
under contract no. 770017. PM acknowledges support from the project PRD 2021 (University of Padova). MS acknowledges funding from the European Union's Horizon 2020 research and innovation programme under the Marie-Sk\l{}odowska-Curie grant agreement No. 794393.
\end{acknowledgements}

\bibliographystyle{aa}
\bibliography{biblio}
\end{document}